\documentclass[12pt,a4paper]{article}
\usepackage[T1]{fontenc}        
\usepackage{lmodern}            
\usepackage[british]{babel}
\usepackage{amsmath}
\usepackage{amssymb}
\usepackage{amsfonts}
\usepackage{setspace}           
\usepackage[dvips]{graphicx}
\usepackage{float}
\restylefloat{figure}
\title{Renormalisation Group Analysis of Turbulent Hydrodynamics}

\author{Dirk Barbi and Gernot M\"unster\\
        Institut f\"ur Theoretische Physik,
        Universit\"at M\"unster\\
        Wilhelm-Klemm-Str.~9, D-48149 M\"unster, Germany\\
        e-mail: munsteg@uni-muenster.de}
\date{version 2, April 28, 2013}
\begin{document}
\maketitle

\begin{abstract}
Turbulent hydrodynamics is characterised by universal scaling
properties of its structure functions. The basic framework for investigations
of these functions has been set by Kolmogorov in 1941. His predictions for the
scaling exponents, however, deviate from the numbers found in experiments
and numerical simulations. It is a challenge for theoretical physics
to derive these deviations on the basis of the Navier-Stokes equations.
The renormalisation group is believed to be a very promising tool for the
analysis of turbulent systems, but a derivation of the scaling properties of
the structure functions has so far not been achieved.
In this work, we recall the problems involved, present an approach in the
framework of the exact renormalisation group to overcome them, and present
first numerical results.\\[5mm]
Turbulent flows - Field-theoretic formulations and renormalisation
\end{abstract}
\section{Introduction}

The theoretical understanding of hydrodynamical turbulence still represents
one of the big challenges of theoretical physics. In his fundamental work
on this subject, Kolmogorov \cite{kolmogorov} introduced structure functions,
describing the moments of velocity differences in a fluid. Assuming
scale-independence in a certain range of distances, he predicted scaling
behaviour of the structure functions, associated with certain classical scaling
exponents. The numbers for the exponents found experimentally and later in
numerical simulations deviate, however, significantly from their classical
values. It is still one of the unsolved problems of classical physics to derive
the scaling behaviour from first principles.

It is generally accepted that the behaviour of an incompressible fluid is on
a fundamental level described by the Navier-Stokes equations, expressing
the conservation of momentum of fluid elements. It should therefore in
principle be possible to deduce the scaling exponents on the basis of the
Navier-Stokes equations. This has, however, not been achieved so far.

A promising approach seems to be the Renormalisation Group (RG), which aims to
describe the dependence of the correlation functions of a given field theory on
the scale on which the system is observed. Beginning with the work of Forster,
Nelson and Stephen \cite{fns}, numerous attempts have been made to apply the
various formulations of the RG to turbulent hydrodynamics, but until today the
observables proposed by Kolmogorov could not be deduced in accordance with
experiment. See e.g.~\cite{dedommartin,aav} for some work on the RG approach to
turbulence.

There are basically two different approaches to the RG, the ``field theoretic'' and the ``exact'' RG. They are related to each other, so that the distinction
may appear artificial, but nevertheless in practice their differences show
up in applications. The field theoretic RG is based on the Callan-Symanzik
equations \cite{callan,symanzik} or closely related approaches, see 
e.g.~\cite{zj}. Under suitable conditions the perturbative calculation of the
RG functions allows to derive the scaling behaviour of correlation functions.
The work of \cite{fns} is based on this approach.

A problem with the application of the field theoretic RG to the study of
turbulence is the fact that it treats only part of the space of all
Hamiltonians, and essentially amounts to an expansion around the case of
laminar flow. Therefore it is not sufficient to capture the essential
properties of the structure
functions. The formulation of the RG most suitable for the study of turbulence
appears to be the Exact Renormalisation Group (ERG) due to Wilson
\cite{wilson,wilsona}, see e.g.~\cite{mccomb}. It explicitly involves the
generating functional of the correlation functions to be studied, and is not
restricted to a small number of couplings. The scaling behaviour to be studied
does not have to be located in the vicinity of the free or laminar theory.
The generating functional can be simplified and
reformulated to suit the analytic methods involved. 
The ERG has been applied to the problem of turbulence by Collina and Tomassini
\cite{coltom}. Based on the Martin, Siggia and Rose functional \cite{msr}, 
they derive a RG flow equation, different from ours, which is studied in an
approximation scheme.

The aim of this work is the following. First, for the generating functional
a functional integral is formulated, which explicitly incorporates all
constraints, and which resolves the constraints and nonlocalities by means of
Lagrange multiplier and auxiliary fields.
The incompressibility condition is implemented in the functional integral,
too. Here we differ from previous work, which in one way
or another omitted the incompressibility condition and/or the pressure term
of the Navier-Stokes equations. 

Then, starting from the action contained
in the functional integral, we formulate a renormalisation group transformation.
The approach presented in this article is based on the ERG. It is especially
helpful, as we shall see, for the analysis of a theory with constraints, like
in our case the incompressibility condition. 
The RG-flow, as we shall discuss in detail, can be understood as the continuous
way of calculating all Feynman graphs of the theory. Keeping this in mind, we
establish a numerical algorithm that calculates the RG-flow by integrating
out the corresponding graphs.
We take advantage of the freedom of choice of a cutoff-function for
the propagator. Our approach leads to rate equations for the RG flow, which
are quite lengthy, but straightforward. They can be iterated quickly
and up to a high number of involved field operators.

In this context we show that the predictions of Kolmogorov can be 
identified as the trivial scaling solutions of this theory. We also show
that non-trivial structures in coupling space exist.
In order to demonstrate the utility of the method, we
have tested the algorithm on well-known theories.
The identification of intermittent exponents in turbulence, however,
has not yet been accomplished due to the numerical complexity of the problem,
and is left for future work.

\section{Basics of Turbulence}

In this section we introduce the basic notions needed in this work, and recall 
some of Kolmogorov's predictions from 1941 (K41).
Reviews can be found for example in \cite{my,frisch,pope,mccomb}. 

\subsection{Navier-Stokes Equations}

The starting point of our considerations are
the full Navier-Stokes Equations (NSE) given by
\begin{equation} \label{nse}
    \partial_{t}v_{\alpha} + v_{\beta} \partial_{\beta}v_{\alpha} - \nu \nabla^{2} v_{\alpha}
  + \frac{1}{\rho}\partial_{\alpha}p = 0,
\end{equation}
where $v$ denotes the $D$-dimensional velocity field (in Navier-Stokes turbulence, $D$ is either $2$ or $3$), 
$\nu$ the kinematic viscosity, $p$ the scalar pressure
field and $\rho$ the density of the fluid.
In incompressible turbulence, the velocity field is required to be
divergence-free,
\begin{equation} \label{inc}
  \mathbf{M}(v):=\partial_{\alpha}v_{\alpha} = 0.
\end{equation}

For fully developed turbulence, statistically homogeneous in space and time,
and statistically isotropic in space, a mechanism is needed to insert energy
into the system, so that an equilibrium flow can develop.  
The standard way of providing this
is to add a stochastic force (stirring force) $f_{\alpha}$ to Eq.~(\ref{nse}) that is long-range correlated:
\begin{equation} \label{nsr}
    \partial_{t}v_{\alpha} + v_{\beta} \partial_{\beta}v_{\alpha} - \nu \nabla^{2} v_{\alpha}
  + \frac{1}{\rho}\partial_{\alpha}p = f_{\alpha}.
\end{equation}
The idea is to bring energy into the flow on large scales, let large structures decay freely
into smaller ones until the energy is finally dissipated into heat (Richardson--cascade).
We model the stochastic force to be Gaussian distributed, with $\delta$-correlation in time,
and a long-range correlation function in space:
\begin{align} 
\langle f(x_1, t) f(x_2, t') \rangle &\equiv F^{-1}(x_1, t, x_2, t) \\
&= - \epsilon \delta(t-t') \nabla^{-2}(x_1, x_2), \label{dissip} 
\end{align}
where $\epsilon$ is the local energy dissipation rate.
$\nabla^{-2}$ denotes the fundamental solution of the Laplacian, e.g.\ in 3 dimensions:
\begin{equation}
\nabla^{-2}(x_1, x_2) = \frac{1}{4\pi|x_1 - x_2|}.\label{longrange}
\end{equation} 
Different forms have been tried for $F$, though in the context of the NSE it is widely believed
that the form of the stochastic force does not influence the statistical characteristics of turbulence.
It should be mentioned, however, that in the case of Burgers turbulence
the intermittent
exponents (to be defined below) clearly depend on the choice of the stirring.

Eq.~(\ref{inc}) is sufficient to eliminate the pressure term, as can be seen in the derivation of the solenoidal NSE as follows.
Operating onto Eq.~(\ref{nsr}) with a divergence operator,
the first and the third terms drop out, as the field is divergence free:
\begin{equation} \label{drucklap} 
\frac{1}{\rho}\nabla^{2}p = \partial_{\beta}f_{\beta} - \partial_{\beta}(v_{\gamma}
\partial_{\gamma}v_{\beta}). 
\end{equation}
Inverting the Laplacian then yields
\begin{equation} 
\frac{1}{\rho}p = \frac{ \partial_{\beta}}{\nabla^{2}}f_{\beta} -
\frac{ \partial_{\beta}}{\nabla^{2}}(v_{\gamma}\partial_{\gamma}v_{\beta}), \end{equation}
which is the above mentioned condition for the pressure field.

One might ask whether the inversion of the Laplacian leads to a unique solution for $p$. Two different solutions
might at best differ by a harmonic function, which is either constant or growing without limits.
The second option is not physical, the first one not relevant as we are only working with
pressure differences.

To obtain the solenoidal NSE, replace
the ``solved'' pressure field into Eq.~(\ref{nsr}):
\begin{eqnarray}\partial_{t}v_{\alpha} - \nu \nabla^{2} v_{\alpha} + \left( \delta_{\alpha \beta}
 - \frac{\partial_{\alpha} \partial_{\beta}}{\nabla^{2}} \right) (v_{\gamma} \partial_{\gamma}
v_{\beta}) &=& \left( \delta_{\alpha \beta} - \frac{\partial_{\alpha} \partial_{\beta}}{\nabla^{2}}
 \right)f_{\beta}\\
  \Leftrightarrow \quad \partial_{t}v_{\alpha} - \nu \nabla^{2} v_{\alpha} + P_{\alpha \beta} (v_{\gamma} \partial_{\gamma}
v_{\beta}) &=&  P_{\alpha \beta}f_{\beta}. \label{nss}
 \end{eqnarray}
From now on we shall investigate the solenoidal NSE. It is important to keep in mind that these are only
equivalent to the full NSE as long as incompressibility is ensured.

Also observe that Eq.~(\ref{nss}) is non-local, as it involves the inverse of the Laplacian operator,
the integral kernel of which is of the form (\ref{longrange}).

The operator $P_{\alpha \beta} = \delta_{\alpha \beta} - 
\frac{\partial_{\alpha} \partial_{\beta}}{\nabla^2}$ is identical to the
transverse projector known from 
electrodynamics.
Due to its appearance the formulae can be
rewritten in a gauge invariant way.
This will facilitate to properly formulate the functional integral discussed below.
As the transverse operator $P$ projects a field
onto its incompressible parts, and observing that the fields we are interested
in are transverse a priori, it is easy to see that
\begin{equation}
P_{\alpha \beta} v_\beta = \left(\delta_{\alpha \beta} - \frac{\partial_\alpha \partial_\beta}{\nabla^2}\right)v_\beta
= \delta_{\alpha \beta}v_\beta = v_\alpha,
\end{equation}
so that we are free to replace $v$ by $Pv$ in Eq.~(\ref{nsr}):
\begin{eqnarray} \label{nsinv}
\mathbf{N}(\vec{v}) &:=& \partial_{t}P_{\alpha \beta}v_{\beta} - \nu \nabla^{2} P_{\alpha \beta}v_{\beta} +
P_{\alpha \beta}(P_{\gamma \delta}v_{\delta}  \partial_{\gamma}P_{\beta \epsilon}v_{\epsilon} ) \nonumber\\
 &=& P_{\alpha \beta}f_{\beta}.
\end{eqnarray}
The resulting equation looks more complicated, but it is invariant under
the same local gauge transformations
\begin{equation} 
v_{\alpha} \rightarrow v_{\alpha} + \partial_{\alpha}\Lambda(\vec{x}) =: U(v), \label{gauge}
\end{equation}
as the vector potential in electrodynamics.
Constraint (\ref{inc}) is still required, but it now acts as a gauge fixing term.

\subsection{Structure Functions and Intermittent Exponents}

In 1941 Kolmogorov introduced a statistical framework for
turbulent hydrodynamics \cite{kolmogorov}. As the theory is Galilean invariant, he proposed that the observables
should be functions of velocity differences, and more specific, he considered the so-called velocity increment
\begin{equation}
v_{\textrm{inc}}(r; x) = (v(r+x) - v(r))\cdot e_x,
\end{equation}
the difference of the velocities at two points separated by a vector $x$, projected onto
the unit vector in $x$-direction.
Suitable observables are the structure functions of order $p$. 
They are defined as the $p$-th moment of the distribution of the absolute value of 
the velocity increment:
\begin{equation} \label{struct}
S_p(x) := \langle |v_{\textrm{inc}}(r; x) |^{p} \rangle_r.
\end{equation}
The average is taken over all spatial points $r$ of a realisation of the turbulent flow.
In the case of homogeneous turbulence, this is supposed to be equivalent to an average
over all histories $v(x, t)$. This average can be defined in terms of a functional integral.

Kolmogorov proposed the existence of a smallest length scale $\lambda$, the
``dissipation scale'', below which physics is no longer dominated by turbulence, but by
dissipation. Dimensional analysis leads to
\begin{equation}
 \lambda = (\nu^3 / \epsilon)^{1/4},
\end{equation}
where $\epsilon$ is the (constant) dissipation rate. Assuming that the turbulent cascade of decaying
vortices happens on scales much larger than $\lambda$, it is argued that
observables don't
depend on it and are thus self-similar, which means power-law functions of the scale:
\begin{equation}
S_p(x) \propto (\epsilon x)^{\xi_p}. \label{scaling}
\end{equation}
By dimensional analysis Kolmogorov deduced
\begin{equation}
\xi_p = \frac{p}{3}.
\end{equation}
It has long been pointed out \cite{ll} that the fundamental assumption, namely the independence of
the smallest scale $\lambda$, is by no means natural, and is in general not fulfilled in
critical systems. This could lead to a scale dependent dissipation rate (or viscosity)
and the breakdown of scaling law (\ref{scaling}). Even though general agreement on this point seems
to be common, the scale dependence could not yet be deduced.

In case that a typical (macroscopic) length scale $L$ can be identified in the system,
the Reynolds number is defined as
\begin{equation}
 \textrm{Re} = \left(\frac{L}{\lambda}\right)^{10/3}. \label{reynolds1}
\end{equation}
$L$ might be the radius of an obstacle of the flow, or, in the context considered here, the correlation length of a choice of the stochastic force. 
Eq.~(\ref{reynolds1}) coincides with 
the more common definition
\begin{equation}
 \textrm{Re} = \frac{L U}{\nu}
\end{equation}
if the typical velocity $U$ is defined to be
\begin{equation}
U = (\epsilon L)^{\frac{1}{3}}.
\end{equation}
 
\section{Generating Functional}

The basic object of the Exact Renormalisation Group (ERG), and many other
field theoretical methods, is the generating functional of correlation
functions. For the case of turbulence, several approaches to define the
generating functional can be found in the literature, see
e.g.~\cite{my,msr,mccomb,lvovproc}. In the work of Martin, Siggia and Rose
\cite{msr} the functional is characterised by means of an infinite hierarchy
of equations, analogous to the field theoretic Dyson-Schwinger equations.
The results of \cite{eg} are obtained in a similar framework.

In order to set up the ERG, it is necessary to formulate the generating
functional in terms of a functional integral. This approach is being
followed e.g.\ in \cite{lvovproc,coltom}.
An apparent problem is that the
incompressibility condition has been disregarded in one way or another.
This condition, however, leads to non-localities which are important for
the correlations in the fluid. In this section
we sketch the derivation of the Martin-Siggia-Rose functional
for the solenoidal NSE, and then show how to respect the incompressibility
condition (\ref{inc}).
As our derivation differs from others in the literature by aspects concerning
the functional determinants and constraints, we prefer to show the line
of arguments in some detail.

\subsection{Fine-Grained Distribution  \label{theta}}

The starting point is the so-called fine-grained probabi\-lity distribution
for the velocity field $v$, obtained by counting all possible solutions to
the NSE.
\begin{equation} 
Z[J] = \int \mathcal{D}v \left< \delta ( v - \mathbf{N}^{-1}(Pf))e^{(v, J)}
\right>_f,
\end{equation}
where $\mathbf{N}$ is defined in Eq.~(\ref{nsinv})\footnote{Here we adopt the
notation of L'vov and Procaccia \cite{lvovproc}.}, and we defined the abbreviation
\begin{equation}
(\mu, A\nu) = \int d^Dx_1 d^Dx_2 dt_1 dt_2
  \ \mu(x_1, t_1) A(x_1, t_1; x_2, t_2) \nu(x_2, t_2). \label{not}
\end{equation}
It is important to notice that the functional integral is an integral over
configurations $v(x,t)$ of the velocity field, representing histories in
space and time, covering the whole range $-\infty < t < \infty$.
The generating functional is thus not a function of physical time $t$.
A few remarks are in order:
\begin{itemize}
\item 
In the functional integral above, $\mathbf{N}^{-1}$ is not to be understood
as the inverse of an operator $\mathbf{N}$ (which might not exist), 
but as a multi-valued operator counting any solution $v$ for
a given realisation of the random force $f$. Observe that the integrand
involves a \emph{functional}
$\delta$-function, meaning that we are searching for histories $v(x, t)$ that
solve the NSE for all $x$ and $t$, rather than a realisation $v(x, t_1)$ at a given time
$t_1$, depending on some initial condition. 
\item  
The average $\langle \cdot \rangle$ is an average over all realisations of $f$,
replacing the spatial average in (\ref{struct}). 
Here we adopt the common assumption that for ho\-mogeneous and isotropic
turbulence, these averages are interchangeable.
This assumption is supported by our results for Burgulence published in
\cite{epl}.
\end{itemize}
Making the average over all realisations of the stochastic force explicit yields
\begin{equation}
Z  \propto \int \mathcal{D}v\int\mathcal{D}f \: \delta ( v -
\mathbf{N}^{-1} (Pf)) e^{-\frac{1}{2}(f, Ff) + (v, J)}. 
\end{equation}
Multiplying the argument of the $\delta$-function by $\mathbf{N}$ leads to a functional determinant that is discussed in detail
in the next paragraph:
\begin{equation}
\Rightarrow Z \propto \int \mathcal{D}v\int\mathcal{D}f \: \delta ( \mathbf{N}
(v) - Pf) e^{-\frac{1}{2}(f, Ff)} 
\det\left\lbrace \frac{\delta \mathbf{N}_{\alpha}(v)(x)}{\delta v_{\beta}(x')} \right\rbrace e^{(v, J)}. 
\end{equation}
The $\delta$-function can be written in terms of a functional Fourier-transformation, introducing an auxiliary field $u$:
\begin{equation}
 \delta(\mathbf{N}(v) - Pf)  \propto  \int \mathcal{D}u \:
e^{i (u, \mathbf{N}(v) - Pf)}.
\end{equation}
We define an action $S_1$ formally by
\begin{equation}
Z = \int \mathcal{D}v \: \mathcal{D}u \: \mathcal{D}f \: e^{-S_1[v,u,f]} \det\left\lbrace \frac{\delta \mathbf{N}_{\alpha}(v)(x)}{\delta v_{\beta}(x')} \right\rbrace e^{(v, J)},
\end{equation}
where the determinant still has to be evaluated. From this
the elements of the Feynman-rules of the theory can be identified.
Let us focus the attention on two parts, which
together lead to the famous $\theta(0)$-problem:
\begin{itemize}
\item $uv$-(diffusion)-propagator:
\begin{figure}[H]
\begin{center}
\includegraphics[width=1.5in]{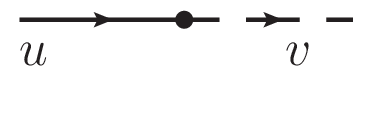}
\end{center}
\end{figure}
\vspace{-1cm}
The corresponding bare two-point-function, also called response function, is proportional to the
Green's function of the diffusion equation, applied to transverse fields:
\begin{equation}
\langle u(x_2, t_2)  v(x_1, t_1)\rangle \propto \frac{1}{ \partial_{t}P_{\alpha \beta} - \nu \nabla^{2} P_{\alpha \beta}}. \label{uvprop}
\end{equation}
In order to ensure causality of the theory, the retarded Green's function
has to be chosen
\begin{equation}
\langle u(x_2, t_2)  v(x_1, t_1)\rangle \propto \theta(t_2 - t_1).
\end{equation}
\item $uvv$-vertex:
\begin{figure}[H]
\begin{center}
\includegraphics[width=1.5in]{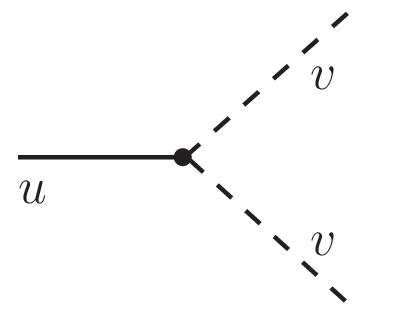}
\label{uvvertex}
\end{center}
\end{figure}
\vspace{-1cm}
This vertex enters the following loop diagram:
\begin{figure}[H]
\begin{center}
\includegraphics[width=2.25in]{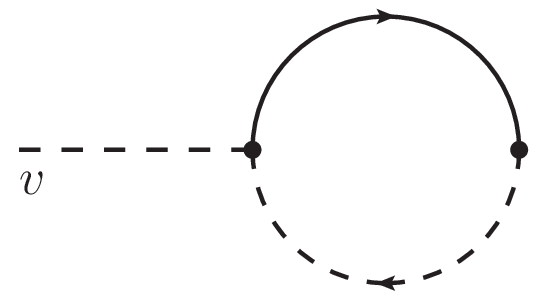}
\caption{\label{uvloop} $u$-$v$-loop of the interaction term}
\end{center}
\end{figure}
As the retarded Green's function is
proportional to $\theta(t_2 - t_1)$, this loop is proportional to
the seemingly ambiguous quantity $\theta(0)$.
\end{itemize}
The appearance of $\theta(0)$ is analogous to the It\={o}-Stratanovich-dilemma,
see e.g.\ \cite{zj}.
In our context, $\theta(0)$ is fixed by the choice of discretisation of
the time derivative inside the functional integral.
For the symmetric (Stratanovich) derivative, one has
$\theta(0) = \frac{1}{2}$,
while in the pure backward (It\={o}) case $\theta(0) = 0$.

To illustrate the contents of the functional integral above,
we integrate out the non-physical fields $f$ and $u$ by means of Gaussian
integration, leading to
\begin{equation}
\label{Zv}
Z = \int \mathcal{D}v \:  \: e^{-S_2[v]} \det\left\lbrace \frac{\delta \mathbf{N}_{\alpha}(v)(x)}{\delta v_{\beta}(x')} \right\rbrace e^{(v, J)},
\end{equation}
with
\begin{equation} S_2[v] = \frac{1}{2} ( \mathbf{N}[v], F \mathbf{N}[v]). \end{equation}
This expression shows that field configurations not solving the NSE are admitted
in the integral, but suppressed by a Gaussian weight.
In principle, Eq.~(\ref{Zv}) can be used as a generating functional, and
all correlation functions can be extracted from it using
functional derivatives. But for the implementation of the RG, it is necessary
to bring the determinant into a suitable form.

\subsection{Functional Determinant}

A straightforward way of writing the functional determinant is
by using anticommuting ghost fields, 
\begin{eqnarray}\label{determinant}
\det\left\lbrace\frac{\delta \mathbf{N}_{\alpha}(v)(x)}{\delta v_{\beta}(y)}\right\rbrace
 &\propto & \int \mathcal{D}\psi^{*} \mathcal{D}\psi \: e^{- i\left( 
\psi^{*}\frac{\delta N}{\delta v}\psi\right)} \\
&=& \int \mathcal{D}\psi^{*} \mathcal{D}\psi \: e^{- i\left(\psi^{*}, \left(
\partial_{t}P- \nu \nabla^{2} \right)\right)\psi +i \left(\psi^{*}, \frac{\delta I}{\delta v}\psi\right)},\label{ghostprop}
\end{eqnarray}
where we dropped a field-independent term and defined $I$ to be the non-linear part of $N$,
\begin{equation}
I_\alpha[v] = v_{\beta} \partial_{\beta}v_{\alpha}.
\end{equation}
This leads to the functional
\begin{equation}
Z = \int \mathcal{D}v \: \mathcal{D}u \: \mathcal{D}f \: \mathcal{D}\psi^* \:\mathcal{D}\psi \:e^{-S[v,u,f,\psi^*, \psi]}  e^{(v, J)},
\end{equation}
where
\begin{equation} \label{wirkgeister}
S[u,v,f,\psi^*, \psi] =
  -i \left( u, \mathbf{N}[v] - Pf \right)  - \frac{1}{2\epsilon}\left(f,\nabla^{2}f\right)
 +i \left(\psi^*, \frac{\delta N}{\delta v} \psi\right).
\end{equation}
It should be noted that the ghost fields, though anticommuting with each other,
can be treated in the numerical procedure calculating the RG flow.
In the algorithm the various contributions are generated according to some
counting scheme.
The terms generated by the ghost fields can be taken fully into account,
order by order, as we have done in several runs of the RG flow equations.

The determinant has a simple graphical interpretation. 
It exactly cancels out the $u$-$v$-loop shown in Fig.~(\ref{uvloop}): 
From Eqs.\ (\ref{uvprop}) and (\ref{ghostprop}), it can be seen
that the $\psi^*\psi$- and the $uv$-propagator are identical. 
It is easily checked that $\psi^*, \frac{\delta N}{\delta v} \psi$ leads to two
terms similar to $u\mathbf{N}[v]$, 
but with $u$ replaced by $\psi^*$ and one $v$-fields replaced by $\psi$. When this vertex is closed to a loop by means 
of a $\psi^*\psi$-propagator, this is numerically identical to graph (\ref{uvloop}).

This greatly simplifies the numerical calculations,
as the program sorts and calculates contributions to the RG-flow according to
their graphical representation. Rather
than simulating two additional fields, and calculating all the graphs, we are thus allowed to drop a certain class of graphs.
The cancellation of certain averages can be proven even non-perturbatively, using the BRS-invariance of the action.

If the functional determinant is expressed in terms of ghost fields, this yields
an extra symmetry, also called BRS-invariance \cite{munoz}.
The action (\ref{wirkgeister}) is indeed invariant under
the infinitesimal transformation
\begin{eqnarray}
\delta u &=& 0,\\
\delta \psi^{*} &=& 0,\\
\delta v &=& \varepsilon \psi^{*}, \\
\delta \psi &=& i\varepsilon u,
\end{eqnarray}
which amounts to ``half a super-symmetry''.
From the Ward identities of this symmetry, the desired result follows on a non-perturbative level:
\begin{equation}
\left\langle \psi \frac{\delta N}{\delta v} \psi^{*} \right\rangle = \left\langle uN\right\rangle.
\end{equation}
The two sides of this equation can be interpreted as the sum of the corresponding graphs discussed in the previous section. 
For details we refer the reader to the explicit proof in \cite{munoz}.

\subsection{Incompressibility Condition}

As has been mentioned before, the incompressibility condition implies
non-localities in the dynamics, which are relevant for the correlations in
the fluid. This becomes apparent by considering models that only differ by the
compressibility condition, and give different statistics.
An obvious example is Burgers' equation, which models fully
compressible fluids and shows bifractal scaling of the structure functions.
It is therefore certainly inadequate to neglect condition (\ref{inc})
completely.

In \cite{lvovproc} the incompressibility condition is considered to be
implied in the functional integration measure. This measure is then, however,
in combination with Gaussian integrands
treated as a functional Gaussian measure, which effectively amounts to
neglecting the incompressibility constraint.

In the context of direct numerical simulations it is sufficient to 
introduce the incompressibility condition through the initial conditions
at time $t=0$. Then the flow stays incompressible without enforcing it by a
particular equation, because in the solenoidal form both the random
force term and the former pressure term lead to incompressible
contributions to the flow. Potential compressible perturbations of a given flow
would die out due to the dissipation term. It is in fact rigorously known
that in two dimensions the statistics of the Navier-Stokes equation
converges to a unique steady state.

This argument does, however, not
apply to our case, since the functional integral represents the equilibrium
statistics and involves configurations
in space and time, i.e.\ it covers the statistics
over whole histories of the fields for all times $-\infty < t < \infty$.
Any possible compressible perturbation of the flow
at a finite time $t$ is going to be amplified
in the negative $t$-direction.
This poses a manifest problem for any numerical approach to the functional
integral due to unavoidable numerical errors.
The slightly compressible flows and the incompressible ones lie dense to
each other in functional space, so that in any numerical application we would
lose control of the boundary conditions completely.

We thus conclude that incompressibility should be taken care of explicitly in
the functional integral. Writing the functional $\delta$-function as
\begin{equation} 
\delta(\partial_{\alpha}v_{\alpha})=\int \mathcal{D}\theta \: e^{i\left(\theta, \partial v\right)},\label{delta}
\end{equation}
would be technically inconvenient in the RG equations.
Therefore the $\delta$-functional is represented in a way familiar from the
initial condition of the kernel of the diffusion equation,
\begin{equation} 
\delta(\partial_{\alpha}v_{\alpha}) \propto
e^{-\frac{1}{2\kappa} \left( \partial_{\alpha}v_{\alpha}, \partial_{\alpha}v_{\alpha}\right) }
\quad \text{in the limit} \quad \kappa \to 0,
\end{equation}
leading to
\begin{equation} \label{functmod1}
Z \propto \int \mathcal{D}u \mathcal{D}v \mathcal{D}f\:
e^{-S[u,v,f]}\det\left\lbrace\frac{\delta \mathbf{N}_{\alpha}(v)(x)}{\delta v_{\beta}(y)}\right\rbrace
\end{equation}
with
\begin{equation} \label{wirkmod1}
S[u,v,f] = 
\frac{1}{2\kappa}(\partial v,\partial v)   -i (u, \mathbf{N}[v] -Pf)
 - \frac{1}{2\epsilon}\left(f, \nabla^{2}f\right),
\end{equation}
where the limit $\kappa \rightarrow 0$ is to be taken when results have been obtained,
in order to enforce incompressibility strictly.
In the functional integral (\ref{functmod1}) only the solenoidal part of the
auxiliary field $u$ is effective, being coupled to the velocity field $v$.
Formally the functional integral implies an integration also over
the longitudinal part of $u$, which would lead to a divergence.
As this integration decouples completely from the remaining degrees of freedom,
it contributes a constant to the generating functional $W[\{J\}]$, discussed in
section 4, and can therefore be neglected.

From now on we shall work with the functional (\ref{wirkmod1}), but formulated
in wavenumber space, which is:
\begin{eqnarray} \label{wirkmod1ft}
S[u,v,f] &=& \int
\left(\frac{d^D p}{(2\pi)^D}\right)  \left(\frac{d^D q}{(2\pi)^D}\right) \delta(p+q) dt \nonumber \\
&& \qquad 
\: \left \lbrace
-\frac{1}{2\kappa}p_{\alpha}v_{\alpha}(p)q_{\alpha}v_{\alpha}(q)
-i u_{\alpha}(p)
\partial_{t}\tilde{P}_{\alpha \beta}(q) v_{\beta}(q) \right. \nonumber \\ && \qquad -i u_{\alpha}(p) \nu
q^{2} \tilde{P}_{\alpha \beta}v_{\beta}(q)   +i
u_{\alpha}(p)\tilde{P}_{\alpha \beta}(q)f_{\beta}(q) \left. +
\frac{1}{2\epsilon}f_{\alpha}(p)q^{2}f_{\alpha}(q) \right\rbrace \nonumber \\& & 
+ \int
\left(\frac{d^D p}{(2\pi)^D}\right)  \left(\frac{d^D q}{(2\pi)^D}\right) \left(\frac{d^D r}{(2\pi)^D}\right)\delta(p+q+ r) dt 
\nonumber \\ && \qquad
u_{\alpha}(p)
\tilde{P}_{\alpha \beta}(p)(\tilde{P}_{\gamma
\delta}(q)v_{\delta}(q)r_{\gamma}\tilde{P}_{\beta
\epsilon}(r)v_{\epsilon}(r)).
\end{eqnarray}

A remark concerning Galilean invariance, as analysed by Hochberg and Berera\cite{bh2007}, is here in order.
The path integral, like the NSE, is invariant under the transformations
\begin{eqnarray}
v(x, t) &=& v'(x', t') + c,\\
x &=& x' + ct,\\
t &=& t'.
\end{eqnarray}
If averages or correlation functions of the field itself are considered, this
would represent a problem that could be overcome by an application of the
Faddeev-Popov method. In our case, however, it does not take effect,
as we only consider averages of velocity differences, which are Galilei
invariant.

\subsection{Non-local Interactions}

The derived action contains interaction terms of the type
\begin{equation}
u_{\alpha}  P_{\alpha \beta}(P_{\gamma \delta}v_{\delta}\partial_{\gamma}P_{\beta \epsilon}v_{\epsilon}), \label{nonlocal}
\end{equation}
which are non-local in coordinate space, but of a very simple form in wavenumber
space. In both cases we need to rewrite (\ref{nonlocal}) in a local way: In
coordinate space,
non-local interactions are at best cumbersome; in wavenumber space, we are going 
to sort the terms
of the action according to their power of momenta, so we try to avoid $\frac{1}{p^2}$-interactions.

We shall proceed in two steps: we will first re-define non-physical fields, and then introduce new fields to
remove the non-locality of interactions. We will end up with a lengthy, but local action that suits our
needs for further analysis.

To make the non-local nature of the interactions more manifest, 
we consider functions in coordinate space within this paragraph.
First we redefine the non-physical fields by
\begin{eqnarray}
u & \rightarrow & \nabla^{2}u , \\
\psi & \rightarrow & \nabla^{2}\psi , \\
\psi^{*} & \rightarrow & \nabla^{2}\psi^{*}.
\end{eqnarray}
Introducing
\begin{equation}
Q_{\alpha \beta} := \nabla^{2}\delta_{\alpha \beta} -
\partial_{\alpha} \partial_{\beta}
\end{equation}
the action is written as
\begin{eqnarray}
-i u_{\alpha}  P_{\alpha \beta}(P_{\gamma \delta}v_{\delta}\partial_{\gamma}P_{\beta \epsilon}v_{\epsilon}) & \rightarrow &
-i  Q_{\alpha \beta} u_{\alpha}\partial_{\gamma}P_{\beta \epsilon}v_{\epsilon}P_{\gamma \delta}v_{\delta}, \label{nl1} \nonumber \\ \\
\psi_{\alpha}^{*}P_{\alpha \beta}(P_{\gamma \delta}\psi_{\delta}\partial_{\gamma}P_{\beta \epsilon}v_{\epsilon}) & \rightarrow &
-\partial_{\beta}Q_{\alpha \gamma}\psi_{\alpha}^{*}Q_{\beta \epsilon}\psi_{\epsilon}P_{\gamma \delta}v_{\delta},\nonumber \\ \\
\psi_{\alpha}^{*}P_{\alpha \beta}(P_{\gamma \delta}v_{\delta}\partial_{\gamma}P_{\beta \epsilon}\psi_{\epsilon}) & \rightarrow &
 Q_{\alpha \beta}\psi_{\alpha}^{*}\partial_{\gamma}Q_{\beta \epsilon}\psi_{\epsilon}P_{\gamma \delta}v_{\delta}.\label{nl2} \nonumber \\
\end{eqnarray}
The functional determinant of these transformations is field independent and can
thus be omitted.

The projector $P$ contains the inverse Laplacian so that
non-local terms of the general form
\begin{equation}
K\frac{1}{\nabla^{2}}L \label{nl3}
\end{equation}
are present. These can be removed by means of new auxiliary fields.
They can be interpreted as transmitting fields that ``carry'' the non-local
interaction from one place to another, thus
replacing it by two local interactions and a propagator. Formally this is
achieved by a Gaussian integral of the type:
\begin{equation}
\int \mathcal{D}\hat{M} e^{-\frac{1}{2} (\hat{M}, \nabla^2 \hat{M})},
\end{equation}
where $\hat{M}$ is the auxiliary field. 
This leads to a new kinetic term $\frac{1}{2} (\hat{M}, \nabla^2 \hat{M})$
in the action, which is independent of all physical fields. 
Shifting the variables as
\begin{equation}
\hat{M} := M + \lambda^{-1} \frac{1}{\nabla^{2}}K + \lambda\frac{1}{2}\frac{1}{\nabla^{2}}L.
\end{equation}
and noticing that
\begin{equation}
-\hat{M}\nabla^{2}\hat{M} + K\frac{1}{\nabla^{2}}L = - M\nabla^{2}M - 2\lambda^{-1}M K - \lambda M L
- \lambda^{-2}K\frac{1}{\nabla^{2}}K - \frac{1}{4}\lambda^{2}L\frac{1}{\nabla^{2}}L
\end{equation}
we get rid of the original, non-local interaction (\ref{nl3}) by replacing it by a new kinetic term for $M$ and new interactions.
Two of them are still non-local, but of the diagonal form
\begin{equation}
\lambda^2L\frac{1}{\nabla^{2}}L.
\end{equation}
They are treated by the same method to get a local action finally.
We add again a Gaussian integral for a new field, say $\hat{N}$, and define
\begin{equation}
\hat{N} := N + \frac{1}{2}\lambda i\frac{1}{\nabla^{2}}L,
\end{equation}
leading to
\begin{equation}
 -\hat{N}\nabla^{2}\hat{N} - \frac{1}{4}\lambda^{2}L\frac{1}{\nabla^{2}}L = -N\nabla^{2}N - i\lambda N L .
\end{equation}
The constant $\lambda$ is needed so that the new fields get a definite dimension.

Applying the method discussed above to action (\ref{wirkmod1ft}), we arrive at the local action
\begin{eqnarray}\label{locaction}
S_{\textrm{loc}} & = &  O_{0}[v, u, f] +
O_{1}[ \phi_{1},\phi_{2}, \phi_{3}, f, v, u] + O_{2}[\phi^{1},\phi_{2}, \phi_{3}, f, v, u],
\end{eqnarray}
with
\begin{align}
O_{0} = &\int
\left(\frac{d^D p}{(2\pi)^D}\right)  \left(\frac{d^D q}{(2\pi)^D}\right) \delta(p+q)\: dt\: \Bigg( -iu_{\alpha}(p)\partial_{t}v_{\alpha}(q) +
iu_{\alpha}(p)f_{\alpha}(q)\Bigg),\nonumber \displaybreak[2]\\
O_{1}  = & \int
\left(\frac{d^D p}{(2\pi)^D}\right)  \left(\frac{d^D q}{(2\pi)^D}\right) \delta(p+q)\: dt \:\Bigg( -
\lambda\phi_{1}(p)q_{\alpha}f_{\alpha}(q) 
+2\lambda^{-1}\phi_{2}(p)
q_{\alpha}u_{\alpha}(q)  \nonumber \\
& \qquad  +2 i \lambda^{-1} \phi_{1}(p)q_{\alpha}u_{\alpha}(q) - i\lambda\phi_{3}(p)q_{\alpha}f_{\alpha}(q)\Bigg) \nonumber \\& 
+ \int \left(\frac{d^D p}{(2\pi)^D}\right)  \left(\frac{d^D q}{(2\pi)^D}\right) \left(\frac{d^D r}{(2\pi)^D}\right)
 \delta(p+q+ r)\:  dt \:
u_{\alpha}(p)v_{\beta}(q)r_{\beta}v_{\alpha}(r), \displaybreak[2]\\ \nonumber \\
O_{2}  = & \int
\left(\frac{d^D p}{(2\pi)^D}\right)  \left(\frac{d^D q}{(2\pi)^D}\right) \delta(p+q)\: dt\:\Bigg(  -i\nu u_{\alpha}(p)q^{2}v_{\alpha}(q)+ \sum_{k=1}^{3}\phi_{k}(p)q^{2}\phi_{k}(q) \nonumber \\ 
& \qquad  + \frac{1}{2\epsilon}f_{\alpha}(p)q^{2}f_{\alpha}(q) \Bigg)\nonumber \\
&+ \int \left(\frac{d^D p}{(2\pi)^D}\right)  \left(\frac{d^D q}{(2\pi)^D}\right) \left(\frac{d^D r}{(2\pi)^D}\right)
 \delta(p+q+ r)\:  dt \: \Bigg(
 -i\lambda p_{\alpha}\phi_{1}(p)v_{\beta}(q)r_{\beta}v_{\alpha}(r)  \nonumber \\
 & \qquad +  \lambda p_{\alpha}\phi_{3}(p)v_{\beta}(q)r_{\beta}v_{\alpha}(r)
\Bigg).
\end{align} 
Here, $\phi_1, \phi_2, \phi_3$ denote the auxiliary scalar fields. 
The terms are sorted according to their order of derivatives.

This is the result for the action $S$. Depending on how the determinant (\ref{determinant}) is expressed, other non-local interactions 
may have to be rewritten in the same way.

\subsubsection{Discussion}

In this section we have shown how to transform non-local into local interactions: either by redefinition of
unphysical fields, or by introduction of intermediate propagators.
A drawback will be that we have to approximate this action to
account for RG-transforma\-tions, and a common way is the derivative expansion. Due to our ``localisation'' of the action, the original terms have been mixed
concerning the order of derivatives. Moreover, the
number of derivatives has increased for most interactions, which means that we would have to expand
the RG-flow to a high order in the derivative expansion. Also, the number of fields involved increases the complexity of the numerical work,
even to the lowest orders.

Nevertheless, this expansion is feasible to any order in derivatives, as shown
in \cite{mydiss} for the first two orders. The results are rather 
lengthy rate equations, which will not be elaborated on here.

\section{Renormalisation Group}

The Exact Renormalisation Group (ERG) originates in the work of
\cite{wilson,wilsona}, based on Kadanoff's block-spin picture
\cite{kadanoff}. For introductions into the theory of the renormalisation
group, see e.g.\ \cite{wk,bg}. It is surprising that some very basic
questions, e.g.\ concerning the renormalisation step and the anomalous
dimension, are still being discussed. Therefore we shall consider this point
in detail in paragraph \ref{sectrenorm}, especially the anomalous dimension
and the graphical representation of the flow.

In this section, we discuss the foundations of the ERG and of the flow
equations. We shall not repeat the derivation of the equations, as this can
be found in a number of articles, but we outline the graphical
representation of the different terms, as it will lay the foundations for
our numerical investigations that closely follow the loop expansion.

\subsection{Form of the Action}

We are looking for a RG-flow of a given theory defined by its generating
functional $Z$. To be definite, let us work with the theory of a vector
field $v_i$, and write $Z$ in the following way:
\begin{eqnarray}
 Z &=& 
\exp\{-W[\{J\}]\}  = \int Dv \exp(-S) \\
&=& \int Dv \exp\left\lbrace -\frac{1}{2}(v_{i},P^{-1}_{v_{ij}} v_{j})
  -(J_{v_{i}},Q^{-1}_{v_{ij}}v_{j})
  -S_{\textrm{int}}[v,\Lambda; \Lambda_0]-S_{0}[J,\Lambda;\Lambda_0]\right\rbrace .
\end{eqnarray}
The action depends on two momentum scales $\Lambda$ and $\Lambda_0$. By
$\Lambda_0$ we denote the scale on which we impose the initial
renormalisation condition - e.g.\ the value of the four-point-function is
fixed to a certain value $\lambda_4$ if all external momenta equal
$\Lambda_0$.

The term $S_{0}$ might look uncommon, but is necessary to pick up terms nonlinear in $J$ that will be
generated by the RG-flow. As initial condition, we set $S_{0}[\Lambda = \Lambda_0] = 0$.

Starting from a renormalised action on scale $\Lambda_0$, the flow is going to generate the renormalised
action on all lower scaled $\Lambda$, which is the second momentum scale involved.  From the RG-perspective,
$S[\Lambda = \Lambda_0]$ plays the role of the initial condition of the flow.

It should be noted that the renormalised action $S[\Lambda]$, also called
Wilsonian effective action, is not identical to the field theoretic
effective action $\Gamma$, which generates the one-particle irreducible
vertex functions. It will contain higher order terms even if the
corresponding 1PI vertex functions vanish.

The flow equations depend on the choice of the kinetic action, so we will define the kinetic term to be
\begin{eqnarray}
S_{\textrm{kin}} &=& \frac{1}{2}(v, P^{-1}v) \\
 &=&  \label{skin}\frac{1}{2(2\pi)^{2D}}\int d^Dp  \; d^Dq  \; \delta(p+q)v(p)P^{-1}\left(\frac{p^2}{\Lambda^2}\right)v(q),
\end{eqnarray}
where we define
\begin{equation} \label{propdef}
  P^{-1}\left(\frac{p^2}{\Lambda^2}\right) = \frac{p^2}{\Lambda^2} C^{-1}\left(\frac{p^2}{\Lambda^2}\right).
\end{equation}
$C$ is the cutoff-function, which has the following properties:
\begin{eqnarray}
C^{-1}\left(\frac{p^2}{\Lambda^2}\right) \rightarrow  1 \quad &\textrm{for}& \quad |p| \rightarrow 0, \label{cprop1}\\
C^{-1}\left(\frac{p^2}{\Lambda^2}\right) \rightarrow  0 \quad &\textrm{for}& \quad |p| \rightarrow \infty, \label{cprop2}\\
C^{-1}(1) &=& \frac{1}{2}.\label{cprop3}
\end{eqnarray}
Though it is by no means necessary, one usually assumes that $C^{-1}$ is monotonous, and that it is a smooth approximation of the 
step function,
thus suppressing degrees of freedom on scales bigger than $\Lambda$, while not effectively altering
those on scales below. The last equation (\ref{cprop3}) is ambiguous, but we define a value for $C^{-1}(1)$, so that the role of 
$\Lambda$ becomes
definite. We will say that the degrees of freedom that are suppressed are ``integrated out'',
as this part of the involved integrals can be interpreted as already being performed.
Apart from the properties (\ref{cprop1}-\ref{cprop3}), we are free
in the definition of $C$. It follows that not even
(\ref{skin}) is enforced; other definitions of the propagator have been tried.
In practice, some propagators will lead to simpler numerical calculations than others.
A very special
choice of $C$ is the sharp cutoff $C_s$:
\begin{eqnarray}
 C_s^{-1}\left(\frac{p^2}{\Lambda^2}\right) = 1 \quad &\textrm{for}& \quad |p| < \Lambda, \label{WHcutoff1}\\
C_s^{-1}\left(\frac{p^2}{\Lambda^2}\right) = 0 \quad &\textrm{for}& \quad |p| \geq \Lambda, \label{WHcutoff2}
\end{eqnarray}
which leads to the Wegner-Houghton equation and will be treated separately.

In a similar matter, we define the following kinetic terms for anti-commuting Grassmann variables:
\begin{equation}
S_{\textrm{kin}} = \frac{1}{2}(\Psi_{\mu},P^{-1}_{\Psi_{\mu\nu}} \Psi_{\nu}), \label{propgrassm}
\end{equation}
where $P^{-1}_{\Psi_{\mu\nu}}$ is an antisymmetric matrix in the indices $\mu$ and $\nu$.

For completeness, we already mention here that we will expand the interaction part of the
action, $S_\textrm{int}$, in powers of the fields to illustrate some examples,
\begin{equation}
 S_\textrm{int} = \sum_k S_{\textrm{int},k},
\end{equation}
where we will call
\begin{equation}
S_{\textrm{int},k} = \int \lambda_k \prod_{i=1}^k \left(d^Dp_i \; v(p_i)\right)\:  \delta(\sum_{j=1}^k p_j) \label{vertex}
\end{equation}
a $k$-vertex. The derivation of the flow equations does not depend on this expansion;
but it is useful in some definite calculations.

\subsection{The Wilson Equation}

\subsubsection{Integrating out degrees of freedom}

The RGE can be derived by calculating the effect of a change of the cutoff
on an action, keeping in mind that both the generating functional and the
correlation functions may not change. A nice derivation of the Wilson-flow
equation is e.g.\ found in \cite{ballthorne}.
\begin{eqnarray}
-\dot{W}&:=&\Lambda\frac{\partial W}{\partial \Lambda} \\
&=& \left\langle \frac{1}{2}(v_{i},\dot{P}^{-1}_{v_{ij}} v_{j})
 +(J_{v_{i}},\dot{Q}^{-1}_{v_{ij}}v_{j})+\dot{S}_{\textrm{int}}+\dot{S}_{0}\right\rangle \\
& = & 0.
\end{eqnarray}
In our case, we will lower the cutoff by lowering $\Lambda$, leaving $\Lambda_0$ as a unit of measurement
unchanged. Here and in the following the dot always denotes the RG flow and
not a derivative with respect to physical time.

Applied to the vector theory, for example, we arrive at the following equation for the interaction term
of the action:
\begin{eqnarray}
\dot{S}_{\textrm{int}} &=& \frac{1}{2}\int_{p}\left\lbrace\frac{\delta S_{\textrm{int}}}{\delta v_{j}}\dot{P}_{v_{ji}}
\frac{\delta S_{\textrm{int}}}{\delta v_{i}}-\frac{\delta}{\delta v_{j}}\dot{P}_{v_{ji}}\frac{\delta S_{\textrm{int}}}{\delta v_{i}}
\right\rbrace,
\label{addrg}
\end{eqnarray}
where we dropped a field-independent term. 
Taking the kinetic term into account, we find the simple equation
\begin{eqnarray}
\dot{S} &=& \dot{S}_{\textrm{int}} + \frac{1}{2}(v_{i}\dot{P}^{-1}_{v_{ij}}v_{j})\\
& = &\frac{1}{2}\int_{p}\left\lbrace\frac{\delta S}{\delta v_{j}}\dot{P}_{v_{ji}}\frac{\delta S}{\delta v_{i}}
-\frac{\delta}{\delta v_{j}}\dot{P}_{v_{ji}}\frac{\delta S}{\delta v_{i}}\right\rbrace - \int_{p} \left\lbrace\frac{\delta S}{\delta v_{i}}\dot{P}_{v_{ik}}P^{-1}_{v_{kj}}v_{j} \right\rbrace. 
\label{rge1}
\end{eqnarray}
The term $\frac{1}{2}\int_{p}\frac{\delta S}{\delta v_{j}}\dot{P}_{v_{ji}}\frac{\delta S}{\delta v_{i}} $ will from now on be
called the link-term of the flow-equation, while we will call $-\frac{1}{2}\int_{p}\frac{\delta}{\delta v_{j}}\dot{P}_{v_{ji}}\frac{\delta S}{\delta v_{i}}$ the loop-term. These names will be justified in the following subsection.
In complete analogy, equations for theories involving Grassmann variables $\psi^*$ and $\psi$ with propagator
(\ref{propgrassm}) can be derived:
\begin{equation}
\dot{S}
= - \frac{1}{2}\int_{p}\left\lbrace\frac{\delta S}{\delta \Psi_{\nu}}\dot{P}_{\Psi_{\nu\mu}}\frac{\delta S}{\delta \Psi_{\mu}}
- \frac{\delta}{\delta \Psi_{\nu}}\dot{P}_{\Psi_{\nu\mu}}\frac{\delta S}{\delta \Psi_{\mu}}\right\rbrace 
+\int_p \left\lbrace \frac{\delta S}{\delta \Psi_{\mu}}\dot{P}_{\Psi_{\mu\lambda}}P^{-1}_{\Psi_{\lambda\nu}}\Psi_{\nu}\right\rbrace . 
\end{equation}
In case of anti-commuting fields it is important to keep track of all extra signs that arise.

These equations describe the lowering of the cutoff, or integrating out of degrees of freedom. 
Before we proceed, we shall discuss the graphical interpretation
of the RG-equations.

\subsubsection{Graphical Representation}

\begin{figure}[H]
\begin{center}
\includegraphics[width=2.25in]{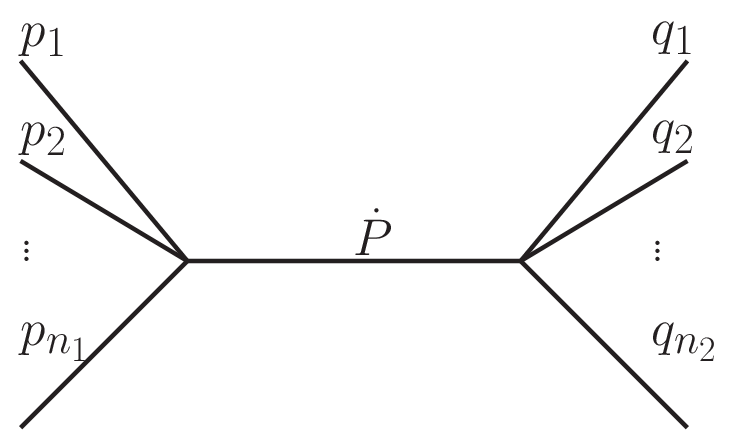} 
\caption{\label{wilslink} Link-term of the Wilson equation in its graphical representation.}
\end{center}
\end{figure}

Let us begin with the interpretation of the link-term. For the time being, we assume that it is applied to a part of the interaction term of the form (\ref{vertex}), a vertex with $n_1+1$ attached lines. Then the functional derivative of this gives us a vertex with $n_1$ lines; the missing line is linked by the part of the propagator
that is integrated out, $\dot{P}$, to a second, similar vertex with, say, $n_2 + 1$ lines. The graphical result is shown in Fig.~\ref{wilslink}. Observe that the functional derivatives automatically lead to the correct symmetry factor of the graph.
\noindent
This graph gives a contribution to the $(n_1 + n_2)$-vertex, proportional to
\begin{equation} \label{lambdalink}
 \int \: \lambda_{n_1 + 1} \lambda_{n_2 + 1} \dot{P}\left(\frac{p^2}{\Lambda^2}\right) \delta(\sum_i p_i + p) \delta(\sum_j q_j - p) d^Dp.
\end{equation}

Since
\begin{equation}
 \delta(\sum_i p_i + p) \delta(\sum_j q_j - p) = \delta(\sum_i p_i + p)\delta(\sum_i p_i + \sum_j q_j),
\end{equation}
one of the two $\delta$-functions just implies the overall conservation of
momentum, and can be eliminated.
The other $\delta$-function will have to be approximated in order to be suitable
for a derivative expansion.

\begin{figure}[H]
\begin{center}
\includegraphics[width=1.75in]{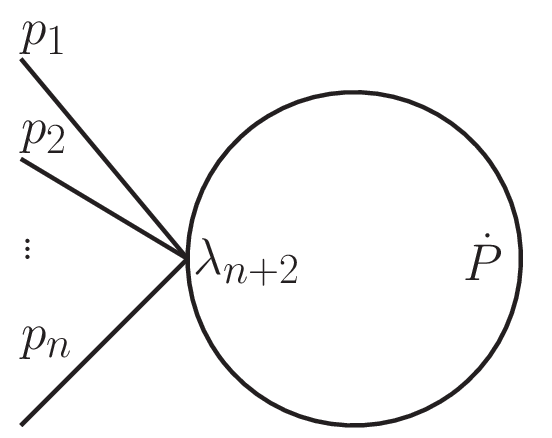}
\caption{\label{wilsloop}Loop calculated in the Wilson equation.}
\end{center}
\end{figure}

The loop-term is equally easy to understand: From a vertex with $n+2$ attached lines, two are joined by a propagator $\dot{P}$ (Fig.~\ref{wilsloop}). Again, the symmetry factor is given correctly. 
\noindent
This graph gives a contribution to the $(n)$-vertex; 
\begin{equation} \label{lambdaloop}
 \int \: \lambda_{n + 2} \dot{P}\left(\frac{p^2}{\Lambda^2}\right) \delta(\sum_{i=1}^{N} p_i) d^Dp.
\end{equation}

So far, we explained the effect of the flow equation as only $S_\textrm{int}$ is concerned. Let us now investigate the contributions of the kinetic term.

\begin{figure}[H]
\begin{center}
\includegraphics[width=1.25in]{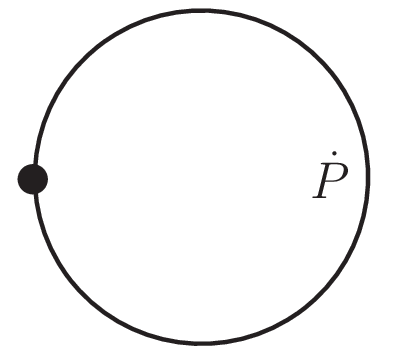}
\caption{\label{wilstrivloop}Field-independent loop that is constructed from the kinetic term of the action.}
\end{center}
\end{figure}

The loop-term generated from the kinetic term (Fig.~\ref{wilstrivloop}) is trivial ,
as it is field-independent and can be dropped.

Let us consider the terms arising when one field derivative in the link-term acts on the interaction, and the other one on the kinetic term; this one is compensated by another term in the RG-eq.:
\begin{equation}
\frac{1}{2}\int_{p}\left\lbrace\frac{\delta S_{\textrm{int}}}{\delta v_{j}}\dot{P}_{v_{ji}}\frac{\delta S_{\textrm{kin}}}{\delta v_{i}}
+\frac{\delta S_{\textrm{kin}}}{\delta v_{j}}\dot{P}_{v_{ji}}\frac{\delta S_{\textrm{int}}}{\delta v_{i}} \right\rbrace
- \int_{p} \left\lbrace\frac{\delta S_{\textrm{int}}}{\delta v_{i}}\dot{P}_{v_{ik}}P^{-1}_{v_{kj}}v_{j} \right\rbrace
= 0.
\end{equation}
The remaining term to be considered is
\begin{equation}
\frac{1}{2}\int_{p}\frac{\delta S_{\textrm{kin}}}{\delta v}\dot{P}\frac{\delta S_{\textrm{kin}}}{\delta v}
- \int_{p} \frac{\delta S_{\textrm{kin}}}{\delta}\dot{P}P^{-1}v
= - \frac{1}{2}\int_{p} v P^{-1}\dot{P}P^{-1}v.
\end{equation}

From
\begin{equation}
0 = - \Lambda \frac{d}{d\Lambda}\left(PP^{-1}\right)
= \dot{P}P^{-1} + P \dot{P}^{-1}
\end{equation}
we get
\begin{equation}
- \frac{1}{2}\int_{p} v P^{-1}\dot{P}P^{-1}v =
\frac{1}{2}\int_{p}v\dot{P}^{-1}v,
\end{equation}
and this is, as defined in (\ref{skin}), the change in the kinetic term.

Let us summarise: We have seen that the RG-flow can be expressed graphically. Iteratively, we
calculate the contributions from all graphs with one propagator $\dot{P}$, and all other propagators $P$, that are inner propagators which have been generated in RG-steps before.
This is simply an application of the product rule:
\begin{equation}
-\Lambda\frac{d}{d\Lambda} \prod_i P\left(\frac{p_i^2}{\Lambda^2}\right) = \sum_i \dot{P}
\left(\frac{p_i^2}{\Lambda^2}\right)\prod_{j\neq i} P\left(\frac{p_j^2}{\Lambda^2}\right).
\end{equation}
Let $\mathcal{G}[f(p)]$ formally denote the sum of all possible Feynman-graphs of
the theory with inner propagators $f(p)$.
The the formal solution to the Wilson equation is
\begin{eqnarray}
 S[\Lambda; \Lambda_0]& =& S[\Lambda = \Lambda_0]  - \int_{\Lambda_0}^{\Lambda} \mathcal{G}\left[\dot{P}(\frac{p^2}{\tilde{\Lambda}^2})\right] \frac{d\tilde{\Lambda}}{\tilde{\Lambda}} \\
 & =& S[\Lambda = \Lambda_0] - \int_{\Lambda}^{\Lambda_0}  \mathcal{G}\left[
 \frac{d}{d\tilde{\Lambda}}P(\frac{p^2}{\tilde{\Lambda}^2})\right] d\tilde{\Lambda} \\
 & =& S[\Lambda = \Lambda_0] -  \mathcal{G}\left[ P(\frac{p^2}{\Lambda_0^2}) - P(\frac{p^2}{\Lambda^2})\right].
\end{eqnarray}
In the limit $\Lambda_0 \rightarrow \infty$ this becomes
\begin{equation}
 S[\Lambda; \Lambda_0] = S[\Lambda = \Lambda_0 = \infty]-  \mathcal{G}\left[ 1-P(\frac{p^2}{\Lambda^2})\right].
\end{equation}
This reveals the meaning of the changes to the action: The RG-flow sums up the part of the propagator
that is cut off iteratively. Notice that we seem to subtract all graphs - this is because we are working with $e^{-S}$ rather than
$e^S$.

Of course the derivation and graphical interpretation of the flow-equation does not apply directly to the
case of a sharp cutoff as defined in (\ref{WHcutoff1}, \ref{WHcutoff2}). As our numerical approach
favours the Wegner-Houghton equation, we discuss it in the following paragraph.

\subsection{The Wegner-Houghton Equation}

For numerical purposes, it is easiest to work with the sharp cutoff function $C_s$ defined in (\ref{WHcutoff1}, \ref{WHcutoff2}).
This changes the form of the flow equation drastically. Again, we shall not
present a derivation
here as it is found in the original literature \cite{wegnerhoughton}, but only
present a short overview. We start from dividing
degrees of freedom into a high-momentum part that is to be integrated out, and
a low-momentum part that is kept. Expressing the integrated part of the functional as a change $\Delta S$ to the action, one finds
\begin{equation} 
e^{-\Delta S} =  \int ' \mathcal{D}v \exp \{ - \int ' L\}, 
\end{equation}
where the prime denotes integration over the momentum shell between $\Lambda-d\Lambda$ and $\Lambda$. 
Expanding the action two second order in the fields gives
\begin{eqnarray}
&&e^{-\Delta S} = \int ' Dv \exp \left\lbrace - \int ' \left\lbrace v \frac{\delta S}{\delta v} + \frac{1}{2}v \frac{\delta^2 S}{\delta v^2}v  \right\rbrace \right\rbrace\\
&=&  \int ' Dv \exp \left\lbrace - \int ' \left\lbrace \frac{1}{2}\left(v \frac{\delta^2 S}{\delta v^2}v + 2 v \frac{\delta S}{\delta v} + \frac{\delta S}{\delta v}
\left(\frac{\delta^2 S}{\delta v^2}\right)^{-1}\frac{\delta S}{\delta v}\right) - \frac{1}{2} \frac{\delta S}{\delta v}
\left(\frac{\delta^2 S}{\delta v^2}\right)^{-1}\frac{\delta S}{\delta v} \right\rbrace \right\rbrace,
\nonumber\\
\end{eqnarray}
where the square has been completed. 
Integrating out the field $v$ in the shell of momentum, which is assumed to be of
thickness $\frac{\Delta\Lambda}{\Lambda} \ll 1$,
we get 
\begin{eqnarray}
e^{-\Delta S} &\propto& \frac{\Delta\Lambda}{\Lambda} \left(\det{\frac{\delta^2 S}{\delta v^2}}\right)^{-\frac{1}{2}} \exp \{ \int '\frac{1}{2} \frac{\delta S}{\delta v}
\left(\frac{\delta^2 S}{\delta v^2}\right)^{-1}\frac{\delta S}{\delta v} \},
\end{eqnarray}
and with the aid of 
\begin{equation} 
(\det A)^{\alpha} = \exp \{ \alpha \textbf{Tr} \ln A \}
\end{equation}
we arrive at the Wegner-Houghton equation
\begin{equation} \label{WH}
\dot{S} = -\frac{1}{2} \int_p \left\lbrace  \frac{\delta S}{\delta v}\frac{\delta S}{\delta v} 
\left(\frac{\delta^2 S}{\delta v^2}\right)^{-1} - \textbf{Tr} \ln \left(\frac{\delta^2 S}{\delta v^2}\right)\right\rbrace. 
\end{equation}
In the derivation it is used that it is sufficient to work in one-loop-order, as higher order
contributions are also of higher order in $d\Lambda$. A proof of this is found in the original work \cite{wegnerhoughton}.

Notice that (\ref{WH}) seems to differ from the Wilson equation~(\ref{rge1}) by an overall-sign; but this is explained as
$\dot{P}$ is negative.

Eq.~(\ref{WH}) is especially convenient for numerical applications, as the contributions to the integrals
can be calculated explicitly. At first glance the logarithm looks problematic,
but it will be shown in the next paragraph that it has a very simple graphical
interpretation, and is thus favourable for our graphically based
program.

\subsubsection{Graphical Representation}

\begin{figure}[H]
\begin{center}
\includegraphics[width=3.25in]{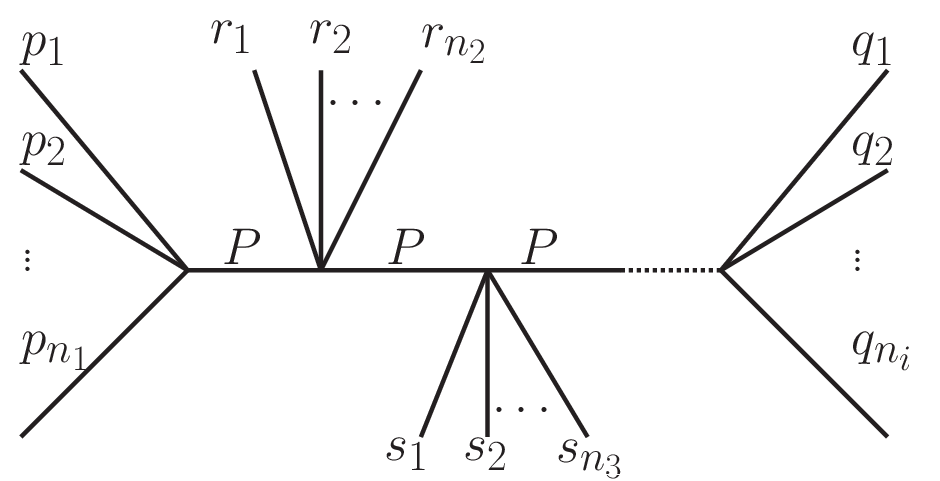}
\caption{\label{whlink} Link to be calculated using the Wegner-Houghton equation. This graph gives a contribution to the vertex with $\sum_i n_i$ outer fields.}
\end{center}
\end{figure}

\begin{figure}[H]
\begin{center}
\includegraphics[width=2.25in]{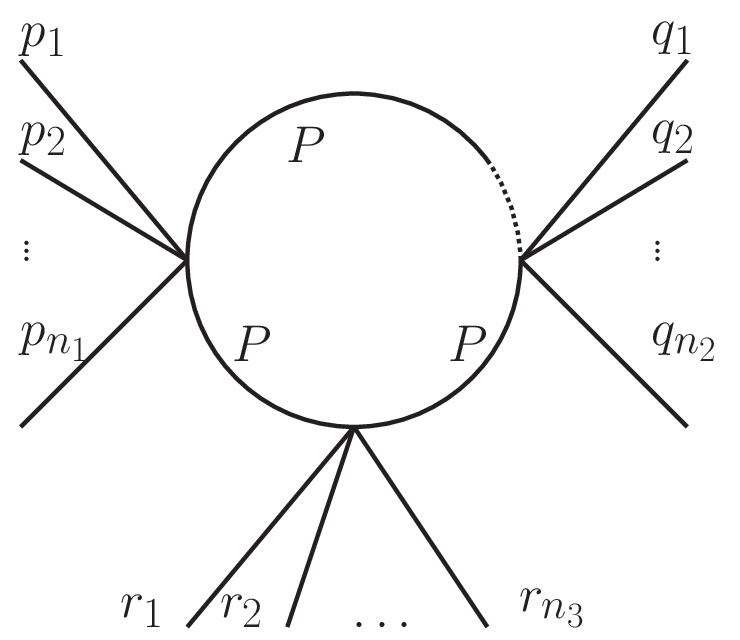}
\caption{\label{whloop} Loop calculated in the Wegner-Houghton equation.}
\end{center}
\end{figure}

For the Wegner-Houghton equation link-term does not involve the
bare propagator alone, but the quantity $\left(\frac{\delta^2 S}{\delta v^2}\right)^{-1}$, which is more than just the inverse
of the kinetic term. Using the geometric series, we can write
\begin{eqnarray}
 &&\left(\frac{\delta^2 S}{\delta v^2}\right)^{-1}
 = \left(\frac{\delta^2 S_{\textrm{kin}}}{\delta v^2}+\frac{\delta^2 S_{\textrm{int}}}{\delta v^2}\right)^{-1}\\
&=& \left(\frac{\delta^2 S_{\textrm{kin}}}{\delta v^2}\right)^{-1} \left(1+\frac{\delta^2 S_{\textrm{int}}}{\delta v^2}\left(\frac{\delta^2 S_{\textrm{kin}}}{\delta v^2}\right)^{-1}\right)^{-1}\\
&=& \left(\frac{\delta^2 S_{\textrm{kin}}}{\delta v^2}\right)^{-1} \sum_i \left\lbrace\frac{\delta^2 S_{\textrm{int}}}{\delta v^2}\left(\frac{\delta^2 S_{\textrm{kin}}}{\delta v^2}\right)^{-1}\right\rbrace^{i}.
\end{eqnarray}
This is the sum of all graphs with $i$ vertices, linked into a line by propagators. The first and the last vertex in the line are also attached to propagators, which link them to terms $\frac{\delta S}{\delta v}$. Again, let us assume first that these act on the interaction part of the action,
thus the chain described above is linked to other vertices. We therefore find the graphical representation Fig.~\ref{whlink}.
In a similar way the graphical representation of the loop-term is derived.
The logarithm is rewritten as:
\begin{eqnarray}
&& \textbf{Tr} \ln \left(\frac{\delta^2 S}{\delta v^2}\right)\nonumber \\ &=&
\textbf{Tr} \ln \left(\frac{\delta^2 S_{\textrm{kin}}}{\delta v^2} + \frac{\delta^2 S_{\textrm{int}}}{\delta v^2}\right) \\
&=& \textbf{Tr} \ln \left(\frac{\delta^2 S_{\textrm{kin}}}{\delta v^2}\left( 1 + \left(\frac{\delta^2 S_{\textrm{kin}}}{\delta v^2}\right)^{-1}
\frac{\delta^2 S_{\textrm{int}}}{\delta v^2}\right)\right)\\
&=& \textbf{Tr} \ln \left(\frac{\delta^2 S_{\textrm{kin}}}{\delta v^2}\right) + \textbf{Tr} \ln \left( 1 + \left(\frac{\delta^2 S_{\textrm{kin}}}{\delta v^2}\right)^{-1}
\frac{\delta^2 S_{\textrm{int}}}{\delta v^2}\right).
\end{eqnarray}
The first term is field-independent and is dropped. The second logarithm is expanded as a Taylor-series, reading
\begin{equation}
\textbf{Tr} \sum_i \frac{(-1)^{i+1}}{i} \left(\left(\frac{\delta^2 S_{\textrm{kin}}}{\delta v^2}\right)^{-1}
\frac{\delta^2 S_{\textrm{int}}}{\delta v^2}\right)^i.
\end{equation}
The corresponding graph is shown in Fig.~\ref{whloop}. It is similar to the link-term before, but closed to a loop by the trace. The factor $\frac{1}{n}$ compensates the rotational symmetry of the graph.
As before, we still have to sort out the link-terms involving the kinetic action. Let us start with an example. 
\begin{figure}[H]
\begin{center}
\includegraphics[width=2.25in]{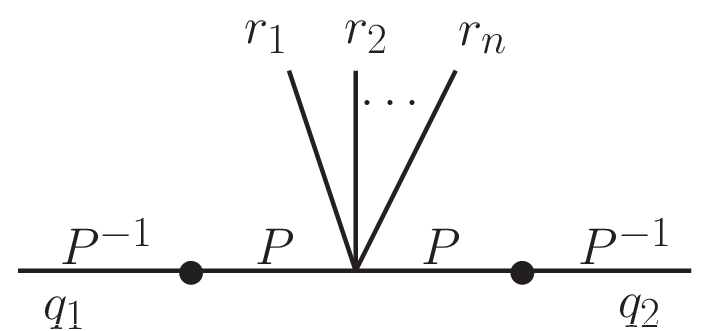}
\caption{\label{whproplink1} Graph of the Wegner-Houghton flow, linking a vertex to two outer propagators in this case.}
\end{center}
\end{figure}
\noindent
The graph in Fig.~\ref{whproplink1} is obviously one of those that arise from the link-term; the reader may focus his attention to one of the $P P^{-1}$ legs. Integration is again over the momentum shell, so 
$P P^{-1} = 1$. The result looks like the vertex, but with the difference that
fields $q_1$ and $q_2$ are depending only on momenta less than $\Lambda - d\Lambda$. These terms can thus be interpreted as integrating out the momenta on
remaining fields. Integrating out more outer fields at the same time would again be of higher order in $d\Lambda$, and can be omitted.

\begin{figure}[H]
\begin{center}
\includegraphics[width=2.25in]{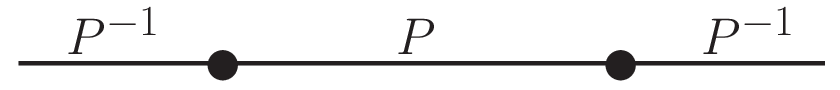}
\caption{\label{whproplink}Graphical representation of the change of the propagator in the Wegner-Houghton flow.}
\end{center}
\end{figure}

The change in the kinetic term itself is again simple, and not even a sign problem arises as in the Wilson-case. The corresponding graph is shown in Fig.~\ref{whproplink},
\noindent
and as $P^{-1} P P^{-1} = P^{-1}$, this is exactly
\begin{equation}
\int' dp \: v P^{-1} v.
\end{equation}
Again, this is precisely the change of the kinetic action, as expected from Eq.~(\ref{skin}).

As in the case of the Wilson equation, we are now able to give a formal solution to the
Wegner-Houghton equation. The final result reads
\begin{equation}
 S[\Lambda; \Lambda_0] = S[\Lambda = \Lambda_0] - \int_{\Lambda}^{\Lambda_0} \mathcal{G}\left[\frac{1}{p^2}\right]d^Dp.
\end{equation}

\subsection{Renormalisation and Rescaling} \label{sectrenorm}

The renormalisation of the field, also called ``field strength renormalisation``
or ``wave function renormalisation'',
is not required in a RG step, but is usually implemented for convenience. As it
is related to the anomalous dimension of the field, it is appropriate to discuss
this point here. We shall demonstrate the concept using
$\phi^4$-theory in $D$-dimensions; for other theories the procedure works in exactly the same way.
Let us emphasise that this step is not
unique to the ERG, but also applied in perturbative renormalisation.

\subsubsection{Field Strength Renormalisation}

\begin{figure}[H]
\begin{center}
\includegraphics[width=2.25in]{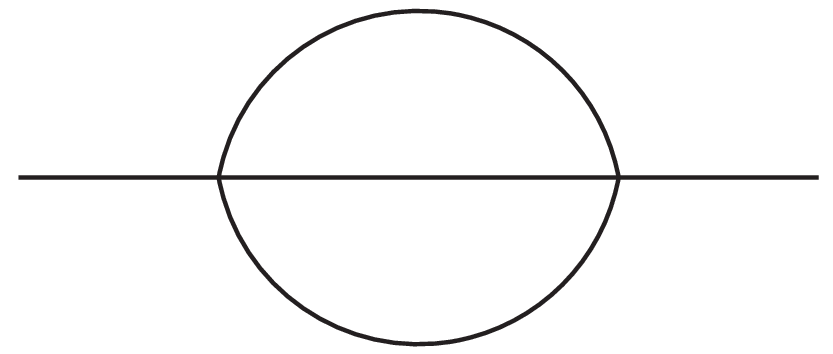}
\caption{\label{sunset} Sunset graph, leading to the simplest contribution to the field strength renormalisation of 
$\phi^4$-theory.}
\label{dh:fig2}
\end{center}
\end{figure}

We started our integration step with the kinetic term
\begin{equation}
S_{\textrm{kin}} = \frac{1}{2}\int \left(\frac{d^Dp}{(2\pi)^D}\right)\left(\frac{d^Dq}{(2\pi)^D}\right) 
C^{-1}\left(\frac{p^2}{\Lambda_0^2}\right)p^2 \: \phi(p) \phi(q) \delta(p+q). 
\label{sorg}
\end{equation}
$S_{\textrm{kin}}$ is defined to be the only term quadratic in fields
and quadratic in the momenta in the limit $p \rightarrow 0$.
After the integration step (lowering the cutoff from $\Lambda_0$ to $\Lambda$), new terms are generated in the interaction part of the action,
that, according to the definition above, should belong to the kinetic term.
Such terms have then to be included in the kinetic term, which changes to some
$S_{\textrm{kin}}'$.
In practice, the first contribution to the kinetic term arises in the second step
of the RG-flow, as it is of two-loop order. 
The simplest graph contributing to field strength renormalisation is the so-called
sunset graph, Fig.~\ref{sunset}.
\noindent
In our case, a graph analogous to Fig.~\ref{sunset} is to be computed by our numerical approach in an iterative way later on,
summing up contributions from every infinitesimal integration.
The result depends on the used renormalisation scheme; by means of
a Taylor-expansion, one can always identify the contribution to the kinetic
action; let us denote it
\begin{equation}
\frac{\eta}{2}\int \left(\frac{d^Dp}{(2\pi)^D}\right)\left(\frac{d^Dq}{(2\pi)^D}\right)
C^{-1}\left(\frac{p^2}{\Lambda^2}\right)p^2 \: \phi(p) \phi(q) \delta(p+q).
\end{equation}
As a renormalisation condition for the field strength, it is commonly required
that the coefficient of the kinetic action in the limit $p\rightarrow 0$ is equal
to $\frac{1}{2}$. According to the definition of
the cutoff-properties, we introduce the field-strength renormalisation factor $Z$ in a way that compensates for the
new term in $S'_{\textrm{kin}}$. If we write for the original action (\ref{sorg})
\begin{equation}
 S_{\textrm{kin}} = Z\frac{1}{2}\int \left(\frac{d^Dp}{(2\pi)^D}\right)\left(\frac{d^Dq}{(2\pi)^D}\right)
C^{-1}\left(\frac{p^2}{\Lambda^2}\right)p^2 \: \phi(p) \phi(q) \delta(p+q),
\end{equation}
we conclude that $Z$ transforms as 
\begin{equation}
\Lambda \frac{\partial}{\partial \Lambda}\ln Z = -\eta. \label{zflow}
\end{equation}
(The sign is negative as in the integration step we actually \emph{lower}
$\Lambda$.)
The initial condition has to be
\begin{equation}
Z[\Lambda = \Lambda_0] = 1, \label{fsr}
\end{equation}
so that (\ref{sorg}) is fulfilled.
Eq.~(\ref{zflow}) is easily integrated to
\begin{equation}
 Z = \left(\frac{\Lambda_0}{\Lambda}\right)^{\eta}.
\end{equation}
$Z$ is now absorbed
into the fields in the following way, which explains the name of field strength renormalisation:
\begin{equation}
 \phi \rightarrow \phi' = \left(\frac{\Lambda_0}{\Lambda}\right)^{\frac{\eta}{2}}\phi \label{anexp}.
\end{equation}
We will use this in the next paragraph to determine the scaling of the field. This is in complete accordance
with renormalisation conditions met in perturbative renormalisation, see for example \cite{pesschroe,collins}.

The kinetic term now reads
\begin{equation}
S_{\textrm{kin}}' = \frac{1}{2}\int \left(\frac{d^Dp}{(2\pi)^D}\right)\left(\frac{d^Dq}{(2\pi)^D}\right)
C^{-1}\left(\frac{p^2}{\Lambda^2}\right)p^2 \: \phi'(p) \phi'(q) \delta(p+q), 
\label{sren}
\end{equation}
which, expressed in renormalised fields, is exactly of the same form as (\ref{sorg}).

It has been stressed by Golner \cite{golner} and Bervillier \cite{bervillier}
that the rescaling step has to be regarded carefully, to account for the
renormalisation step consistently.

\subsubsection{Rescaling}

The last step in the renormalisation group process is the rescaling of the momenta, and the
functions thereof. Define new momenta $\tilde{p}$ by
\begin{equation}
\tilde{p} = \left(\frac{\Lambda_0}{\Lambda}\right) p.
\end{equation}
The replacement
\begin{equation}
p \rightarrow \left(\frac{\Lambda}{\Lambda_0}\right) \tilde{p}
\end{equation}
changes the cutoff function in the expected way:
\begin{equation}
C^{-1}\left(\frac{p^2}{\Lambda^2}\right) \rightarrow  \tilde{C}^{-1}\left(\frac{\tilde{p}^2}{\Lambda_0^2}\right),
\end{equation}
so we are ready to identify the new with the old cutoff.

From the renormalisation step, it is now easy to deduce the scaling of a field. Beginning with the original kinetic term (\ref{sorg}), we conclude that, as
$S_{\textrm{kin}}$ does not scale at all, 
the rescaled field $\tilde{\phi}$ has to be scaled as
\begin{equation}
 \tilde{\phi}(\tilde{p}) = \left(\frac{\Lambda_0}{\Lambda}\right)^{\frac{-D-2}{2}} \phi(p),
\end{equation}
where the canonical dimension $D_{\phi, \textrm{can}} = \frac{D-2}{2}$ appears.
Eq.~(\ref{anexp}) then fixes the anomalous exponent:
\begin{equation}
 \tilde{\phi'}(\tilde{p}) = \left(\frac{\Lambda_0}{\Lambda}\right)^{-\frac{D +2 - \eta}{2}}\phi(p).
\end{equation}
One can now now see that due to the effect of the rescaling of the cutoff-function, the renormalised
kinetic action indeed does not scale. As the cutoff is a function of the ratio $p^2 / \Lambda^2$,
a change in $p$ has the inverse effect as the same change in $\Lambda$. By this, we reverse the
effect of the integration step. As a part of this, terms are re-distributed back to the interaction of
the theory.
Now, in the kinetic term the canonical scaling of the fields is compensated
by the integration measure, and the anomalous scaling by the renormalisation step,
so that finally
\begin{equation}
\tilde{S}_{\textrm{kin}}' = \frac{1}{2}\int \left(\frac{d^D\tilde{p}}{(2\pi)^D}\right)\left(\frac{d^D\tilde{q}}{(2\pi)^D}\right)
\tilde{C}^{-1}\left(\frac{\tilde{p}^2}{\Lambda_0^2}\right)\tilde{p}^2 \: \tilde{\phi'}(\tilde{p}) \tilde{\phi'}(\tilde{q}) \delta(\tilde{p}+\tilde{q}), 
\label{sresc}
\end{equation}
is identical to (\ref{sorg}) as a function of the rescaled quantities, as desired. We will drop the tildes and primes, formally getting back to (\ref{sorg}).

\subsection{The Interaction Terms}

The steps discussed for the kinetic terms have to be
applied to the interaction terms, too. As an example, let us consider
the four-field-interaction
\begin{equation}
\int d^D p_1\:  \int d^D p_2\: \int d^D p_3\: \lambda_4 
\phi(p_1)\phi(p_2)\phi(p_3)\phi(p_4) \delta(p_1 +p_2 +p_3 + p_4).
\end{equation}

\subsubsection{Renormalisation}

When $\phi$ is changed to the renormalised field $\phi'$, without changing the interaction term, the coupling has to be renormalised as follows:
\begin{eqnarray*}
 && \int d^D p_1\:  \int d^D p_2\: \int d^D p_3\:  \lambda_4 \phi(p_1)\phi(p_2)\phi(p_3)\phi(-p_1-p_2-p_3) \\
&& \rightarrow \int d^D p_1\:  \int d^D p_2\: \int d^D p_3\: \lambda_4 \left(\frac{\Lambda_0}{\Lambda}\right)^{\frac{-4\eta}{2}} \phi'(p_1)\phi'(p_2)\phi'(p_3)\phi'(-p_1-p_2-p_3)\\
&& \Rightarrow \lambda_4'= \lambda_4 \left(\frac{\Lambda_0}{\Lambda}\right)^{\frac{-4\eta}{2}}.\\
&& \Leftrightarrow \lambda_4 \rightarrow \left(\frac{\Lambda_0}{\Lambda}\right)^{\frac{-4\eta}{2}}\lambda_4'.
\end{eqnarray*}
For an infinitesimal integration step this amounts to
\begin{equation}
 \dot{\lambda}_4 = -4\,\frac{\eta}{2}\lambda_4.
\end{equation}
For a general interaction $S_\textrm{int}$ with any number of vertices,
this generalises to
\begin{equation}
 \dot{S}_{\textrm{int, Ren}} = \frac{\eta}{2}\int\phi\frac{\delta S_{\textrm{int}}}{\delta \phi}, \label{sintren}
\end{equation}
as the operator $\int\phi\frac{\delta}{\delta \phi}$ counts the number of fields in a vertex.

\subsubsection{Rescaling}

The contributions from the rescaling step are:
\begin{itemize}
 \item Integral: For each integration measure, we get a factor $D$, and there is one integration measure less than
	there are fields (because of the $\delta$-function), so we get a contribution
	\begin{equation}
	 \dot{S}_{\textrm{int}, dp} = - D \int\phi\frac{\delta S_{\textrm{int}}}{\delta \phi} +D S_{\textrm{int}}
	\end{equation}
 \item Momentum: The vertex will depend explicitly on the momentum, so we
introduce another operator
	$\int\phi(p) p \left(\frac{\partial}{\partial p}\right)' \frac{\delta}{\delta \phi(p)}$ that counts the powers of momenta in each
	vertex. The prime at the derivative indicates that it is not acting upon the momentum conserving $\delta$-function. We get the contribution
	\begin{equation}
	 \dot{S}_{\textrm{int}, p} = - \int\phi(p) p \left(\frac{\partial}{\partial p}\right)' \frac{\delta S_{\textrm{int}}}{\delta \phi(p)}.
	\end{equation}
 \item Fields: As derived above, each field brings a contribution proportional to $-\frac{D +2 - \eta}{2}$, so in total we
	find
	\begin{equation}
	 \dot{S}_{\textrm{int}, \phi} = \frac{D +2 - \eta}{2} \int\phi\frac{\delta S_{\textrm{int}}}{\delta \phi}.
	\end{equation}
 \item Renormalised Coupling: any coupling is renormalised according to
(\ref{sintren}), so it scales itself anomalously,
	exactly compensating the anomalous scaling of the fields:
	\begin{equation}
	 \dot{S}_{\textrm{int}, \lambda} =  \frac{\eta}{2} \int\phi\frac{\delta S_{\textrm{int}}}{\delta \phi}.
	\end{equation}
\end{itemize}
Summing up all contributions yields the rescaling term
\begin{equation}
 \dot{S}_{\textrm{int, Rescaling}} = D S_{\textrm{int}} - \int\phi\left(\frac{D - 2}{2} +  p \left(\frac{\partial}{\partial p}\right)'
\right)\frac{\delta S_{\textrm{int}}}{\delta \phi}.
\end{equation}

\subsection{The RG-Equation}

From the previous discussion, the resulting flow equation for the interaction
term is
\begin{eqnarray}
\dot{S}_{\textrm{int}} =&& \frac{1}{2}\int_{p}\left\lbrace\frac{\delta S_{\textrm{int}}}{\delta v_{j}}\dot{P}_{v_{ji}}
\frac{\delta S_{\textrm{int}}}{\delta v_{i}}-\frac{\delta}{\delta v_{j}}\dot{P}_{v_{ji}}\frac{\delta S_{\textrm{int}}}{\delta v_{i}}
\right\rbrace \nonumber \\
&& - \int\phi\left(\frac{D - 2 -\eta}{2} +  p \left(\frac{\partial}{\partial p}\right)'
\right)\frac{\delta S_{\textrm{int}}}{\delta \phi}
+ D S_{\textrm{int}}.
\label{stand}
\end{eqnarray}
This equation depends on the choice of propagator (\ref{propdef}). 
As an example consider the inverse propagator to be given by
\begin{equation}
 \tilde{P}^{-1} = \Lambda^k C^{-1}\left(\frac{p^2}{\Lambda^2}\right), \quad k=2, \label{alterprop}
\end{equation}
as often found in literature. Then the resulting equation is the one given
by Bervillier \cite{bervillier} or Golner \cite{golner}:
\begin{eqnarray}
\dot{S}_{\textrm{int}} =&& \frac{1}{2}\int_{p}\left\lbrace\frac{\delta S_{\textrm{int}}}{\delta v_{j}}\dot{\tilde{P}}_{v_{ji}}
\frac{\delta S_{\textrm{int}}}{\delta v_{i}}-\frac{\delta}{\delta v_{j}}\dot{\tilde{P}}_{v_{ji}}\frac{\delta S_{\textrm{int}}}{\delta v_{i}}
\right\rbrace \nonumber \\
&&- \int\phi\left(\frac{D + 2 -\eta}{2} +  p \left(\frac{\partial}{\partial p}\right)'
\right)\frac{\delta S_{\textrm{int}}}{\delta \phi}
+ D S_{\textrm{int}}, 
\label{golner}
\end{eqnarray}
which in turn is equivalent to Wilson's equation.

Let us point out that both equations (\ref{stand}) and (\ref{golner}) are
correct, even though they seem to differ by a sign. 
The equivalence is obscured by a different choice of propagator functions, taking
advantage of the reparameterisation-invariance of the equation.

The propagator (\ref{alterprop}) indeed has some advantages, as in principle
other values for $k$ are also possible, and simplify the implementation of K41,
as we shall see. On the other hand, the derivative of (\ref{alterprop}) is
more complicated and is especially inconvenient if the
derivative expansion is applied.

The complete RG-equation, including vectorial and Grassmannian fields,
finally reads
\begin{eqnarray}
\dot{S}
& = &\frac{1}{2}\int_{p} \dot{P}_v \left\lbrace\frac{\delta S}{\delta v_{i}}\frac{\delta S}{\delta v_{i}}-\frac{\delta}{\delta v_{i}}
\frac{\delta S}{\delta v_{i}} -2P^{-1}_v\frac{\delta S}{\delta v_{i}}v_{i}\right\rbrace \nonumber \\
& & + \int_{p} \dot{P}_\Psi  \left\lbrace \frac{\delta S}{\delta \psi^{*}_{i}}\frac{\delta S}{\delta \psi_{i}}
- \frac{\delta}{\delta \psi^{*}_{i}}\frac{\delta S}{\delta \psi_{i}} + P^{-1}_\Psi
\left(\frac{\delta S}{\delta \psi^{*}_{i}} \psi^{*}_{i}+\frac{\delta S}{\delta \psi_{i}} \psi_{i}\right)\right\rbrace \nonumber \\
& & -  (D+D_{v_{i},\textrm{kan}} - \frac{\eta_{v_{i}}}{2}) \int_{p} v_{i}\frac{\delta S}{\delta v_{i}}
- (D+ D_{\psi^{*}_{i},\textrm{kan}} - \frac{\eta_{\psi^{*}_{i}}}{2})\int_{p}\psi^{*}_{i}\frac{\delta S}{\delta \psi^{*}_{i}} \nonumber \\
& & - (D+ D_{\psi_{i},\textrm{kan}} - \frac{\eta_{\psi_{i}}}{2})\int_{p}\psi_{i}\frac{\delta S}{\delta \psi_{i}}\nonumber \\
& & -  \int_{p} v_{i}p\frac{\partial '}{\partial p}\frac{\delta S}{\delta v_{i}} - \int_{p}\psi^{*}_{i}p
\frac{\partial '}{\partial p}\frac{\delta S}{\delta \psi^{*}_{i}} - \int_{p}\psi_{i}p\frac{\partial '}{\partial p}
\frac{\delta S}{\delta \psi_{i}} \nonumber \\
& & +DS. \label{rg5} \end{eqnarray}
This equation is general enough to cover our intended applications, including
different ways of considering the functional determinant Eq.~(\ref{determinant}).

\section{Derivative expansion}

The actions involved in the RG flow represent infinitely many degrees
of freedom and have to be approximated in the context of numerical investigations.
A common way of approximation is the derivative expansion, see e.g.\ \cite{hasenfratz} and \cite{morris}.
Applied to the scalar theory, it amounts to expanding the action in powers of derivatives:
\begin{equation}
 S = \frac{1}{2}\int_x\; Z(\phi(x))\left(\partial \phi\right)^2 + V(\phi(x)) + \mathcal{O}(\partial^4),
\end{equation}
in contrast to an expansion in powers of fields,
which can be seen as expansion around a weak field. As a special case, in the Local Potential Approximation (LPA)
the action is reduced to an interaction term depending only locally on the field $\phi(x)$ (and not on its derivatives)
and a kinetic term whose coefficient $Z$ is held constant throughout the flow:
\begin{equation}
 S_{\textrm{LPA}} = \frac{1}{2}\int_x\left(\partial \phi\right)^2 + V(\phi(x)).
\end{equation}
Applying the loop- and link-terms to the action expanded in powers of fields
leads to rate equations
for the coefficients. Let us, as an example, apply the Local Potential Approximation (LPA) to Eqs.~(\ref{lambdalink}) and (\ref{lambdaloop}),
expanded in powers of fields.
Starting with Eq.~(\ref{lambdalink}) for the link-term, the non-trivial part of
graph \ref{wilslink} is proportional to
\begin{eqnarray}
\dot{\lambda}_{n+m} &\propto& \int \dot{P}\left(\frac{p^2}{\Lambda^2}\right) \lambda_{n+1}(p_1, \ldots, p_n, p)\lambda_{m+1}(q_1, \ldots, q_m, p)
\nonumber \\
&& \qquad \qquad \times 
\delta(p_1 + p_2 + \ldots + p_n + p) \delta(q_1 + q_2 + \ldots + q_m - p) d^Dp.
\nonumber\\
&=&
\dot{P}\left(\frac{\sum_i p_i^2}{\Lambda^2}\right) \lambda_{n+1}(p_1, \ldots, p_n, -\sum_i p_i)\lambda_{m+1}(q_1, \ldots, q_m, \sum_j q_j)
\delta(\sum_i p_i + \sum_j q_j).
\nonumber\\
\end{eqnarray}
In the LPA, the couplings are approximated to be momentum-independent, and
developing the cutoff to zeroth order in the momenta gives
\begin{equation}
\dot{\lambda}_{n+m, \textrm{LPA}} = \lim_{p \rightarrow 0}\dot{P}\left(\frac{\sum_i p_i^2}{\Lambda^2}\right) \lambda_{n+1, \textrm{LPA}}\lambda_{m+1, \textrm{LPA}}.
\end{equation}
A difficulty is the momentum-dependence of the factor $\dot{P}\left(\frac{\sum_i p_i^2}{\Lambda^2}\right)$ which we need to expand, according to the derivative expansion. The result obviously depends on the choice of the cutoff; if we apply it to the LPA, we can subsume the result into the constant
\begin{equation} 
\tilde{P}_1 := \lim_{p\rightarrow 0}\dot{P}\left(\frac{\sum_i p_i^2}{\Lambda^2}\right).
\end{equation}
If the cutoff is an approximation of the step function, the limit is expected
to converge, and $\tilde{P}_1 = 0$. This clearly is not an option, as it would suppress the non-trivial character of the RG-flow.

On the other hand, the loop-equation (\ref{lambdaloop}) leads in the LPA to
\begin{eqnarray}
\dot{\lambda}_{n} &=& \int \dot{P}\left(\frac{p^2}{\Lambda^2}\right) \lambda_{n+2}(p_1, \ldots, p_n, p,- p)\delta(p_1 + p_2 + \ldots + p_n)d^Dp \\ \Rightarrow \dot{\lambda}_{n, \textrm{LPA}}
&=& \lambda_{n+2, \textrm{LPA}}\int \dot{P}\left(\frac{p^2}{\Lambda^2}\right)d^Dp \\
&=& \lambda_{n+2, \textrm{LPA}}\Omega_{D-1} \int \left(\frac{d}{dp} P\left(\frac{p^2}{\Lambda^2}\right)\right) p^D dp. 
\end{eqnarray}
Again, the integral depends on the choice of the cutoff; for the LPA we write
\begin{equation}
\Omega_{D-1}\int \left(\frac{d}{dp} P\left(\frac{p^2}{\Lambda^2}\right)\right) p^D dp \rightarrow \tilde{P}_0.
\end{equation}
Rather than to specify a cutoff, in the LPA it is sufficient to define the constants $\tilde{P}_1$ and $\tilde{P}_0$. In higher orders of the
derivative expansion, additional information concerning the cutoff will be required.

In the case of a vector theory in three dimensions, the situation is not that simple, as products of
the type $v_i v_i$ or any contraction with other three-component
fields will be present. We need to keep track of this to calculate the contributions to a renormalisation group flow,
so we propose to expand the terms of the action in powers of fields and momenta in the following way:
\begin{equation}
\begin{split} V = &\sum_{x,r,q,A}\quad  _{(x_1, x_2, x_3)}^{\hspace{3.3em} q}V_{(r_1, r_2, \ldots r_6)}^{(A_1, A_2,\ldots A_6)}\\ & \times
(vv)^{A_1}(uu)^{A_2}(ff)^{A_3}(vu)^{A_4}(vf)^{A_5}(uf)^{A_6}\\
& \times
(v\psi^{*})^{r_1}(v\psi)^{r_2}(u\psi^{*})^{r_3}(u\psi)^{r_4}(f\psi^{*})^{r_5}(f\psi)^{r_6}(\psi^{*}\psi)^{q}\\
& \times
(\phi^1)^{x_1}(\phi^2)^{x_2}(\phi^3)^{x_3}.
\end{split}
\end{equation}
From now on, we will work with the coefficients 
\begin{equation} _{(x_1, x_2, x_3)}^{\hspace{3.3em} q}
V_{(r_1, r_2, \ldots r_6)}^{(A_1, A_2,\ldots
 A_6)}.\end{equation}
In first order, the terms of  the derivative expansion are even more complicated, as we also have to keep
track of terms like $p_i v_i(q)$.

As the overall number of momenta is fixed for each term, and the action itself is
scalar, we get the following
possible values for the indices of $V$:
\begin{xalignat}{2}
 x_i &\in \mathbf{N_{0}}, & \quad A_i &\in
 \mathbf{N_{0}},\\
 r_i &\in \lbrace 0,1 \rbrace, & \quad q &\in \lbrace 0, \ldots, D - r_1 -r_3 -r_5\rbrace , \\
 &&\quad r_1 + r_3 + r_5 &=
 r_2+ r_4 + r_6,
\end{xalignat}
and for $Z$ equivalently. 

\section{Application to Turbulence}

Applying the RG to turbulence, a point of central importance is to specify
how the RG transformations should act on the degrees of freedom contained
in the action. In the case at hand we decide to consider transformations
that describe pure spatial rescalings, while physical times are not being rescaled.
In the language of the block spin RG, this represents a block spin
transformation highly anisotropic in the coordinates $(t, x)$, in which the
blocking is applied to the three spatial coordinates $x$ only.
In Fourier space, the RG transformation acts on three-dimensional momenta,
but not on frequencies.
The reason for this approach is twofold. First, the goal of the
RG calculations is to study the scaling behaviour of the structure
functions, which are spatial correlation functions and do not involve
physical time $t$. The RG transformations relevant for this are spatial
ones. Secondly, this allows to
apply the formalism discussed in the previous section without fundamental
modifications, because pure spatial scalings are being considered there. As
a consequence, the loop integrals contributing to the flow equations are
momentum space integrals and do not involve frequencies.

The complete correlation functions of a given theory do of course
not depend on how the action is divided into a kinetic part and 
interaction terms. 
In order to implement RG transformations it is, however,
crucial to specify the kinetic part of the action, because it contains
the cutoff-function, which is the primary source of the dependence 
of the action on the cutoff $\Lambda$.
The kinetic part appropriate for the kind of RG transformations intended
here, consists of the terms quadratic in the fields and in the spatial
derivatives in the action corresponding to the Navier-Stokes equation,
Eqs.~(\ref{wirkmod1},\ref{wirkmod1ft}).
Consequently, terms linear in $\partial_t$ are treated as parts of the
interaction. The RG transformations will thus involve momentum/space integrals
but not frequency/time integrals. This does, however, not mean that
the time dependence of the theory is eliminated; it just does not enter
the integrals effecting the RG transformations.

Fields with time derivatives are to be tracked in the book
keeping as they will be generated by the RG flow. We denote
the number of time
derivatives in a term by \textit{Der}.
In the derivative expansion the coefficients are correspondingly labelled
\begin{equation} \label{coefficients}
 _{(x)}^{ \; \;q}V_{(r)}^{(A)}[\mathit{Der}].
\end{equation}

For the final assembly of the rate equations, we need the scaling dimensions
and exponents $k$ for the involved fields. The canonical dimension for the
velocity field $v$ is derived from the energy flow $\Pi'$, see
\cite{frisch}:
\begin{equation} \Pi'_{l} \propto \frac{\sqrt{\left\langle v^2(l)\right\rangle^{3}}}{l} \propto \epsilon,
\end{equation}
from which we see that $\sqrt{\left\langle v^2(l)\right\rangle} \propto l^{
\frac{1}{3}}$. In wavenumber space this implies
\begin{alignat}{2}
\left[\partial_{t}\right]&= \frac{2}{3},& \qquad \left[\nu\right]&= -\frac{4}{3},\\
\left[v\right]&=-D-\frac{1}{3}.&&
\end{alignat}
The scaling dimensions of the non-physical fields and constants are then:
\begin{alignat}{2}
\left[f\right]&= -D+\frac{1}{3},& \qquad \left[u\right]&= -\frac{1}{3},\\
 \left[\rho\right]&= -D + \frac{8}{3}, &\qquad \left[\psi^{*}\right]&=\left[\psi\right]= -\frac{D}{2} - \frac{1}{3}, \\
\left[\lambda\right]&= \frac{D}{2}-\frac{1}{3},&
\left[\phi^{i}\right]&= -\frac{D}{2}-1 \quad \forall i,
\end{alignat}
and the exponents $k$:
\begin{alignat}{2}
k_v &= D + \frac{2}{3},& \qquad k_{\phi^{i}} &= 2 \quad \forall i,\\
k_{f} &= D-\frac{2}{3},& \qquad k_{u}&= -D+\frac{2}{3},\\ 
k_{\Psi} &= \frac{2}{3}. &&
\end{alignat}
In this way we arrive at the rate equations that we simulated numerically.
These equations are quite lengthy, details are presented in \cite{mydiss}.
Here we only present the LPA:
\begin{align} \label{rateeq}
- \Lambda\frac{d}{d\Lambda}\left( _{(x)}^{
 \; \; q}V_{(r)}^{(A)}[\mathit{Der}]\right)  =
 &\frac{-D-\frac{2}{3}+\eta_v}{2}\left( \tilde{P}_{v, 1}\left( \frac{ \partial V}{\partial v} \frac{\partial V}{\partial v}\right)_{q, (r)}^{(x), (A)} - \tilde{P}_{v, 0}\left( \frac{ \partial^2 V}{\partial v^2}\right)_{q, (r)}^{(x), (A)}
 \right)\nonumber\\
&+ \frac{D -\frac{2}{3}+\eta_u}{2}
\left( \tilde{P}_{u, 1}\left( \frac{ \partial V}{\partial u} \frac{\partial V}{\partial u}\right)_{q, (r)}^{(x), (A)} - \tilde{P}_{u, 0}\left( \frac{ \partial^2 V}{\partial u^2}\right)_{q, (r)}^{(x), (A)}\right)\nonumber\\
 &+\frac{-D+\frac{2}{3}+\eta_f}{2}\left(\tilde{P}_{f, 1}\left( \frac{ \partial V}{\partial f} \frac{\partial V}{\partial f}\right)_{q, (r)}^{(x), (A)} - \tilde{P}_{f, 0}\left( \frac{ \partial^2 V}{\partial f^2}\right)_{q, (r)}^{(x), (A)} \right)\nonumber\\
 &+ \left(-\frac{2}{3}+\eta_{\Psi}\right)\left(\tilde{P}_{\Psi, 1}
 \left( \frac{ \partial V}{\partial \psi^*} \frac{\partial V}{\partial \psi}\right)_{q, (r)}^{(x), (A)} - \tilde{P}_{\Psi, 0} \left( \frac{ \partial^2 V}{\partial \psi^*\psi}\right)_{q, (r)}^{(x), (A)}
 \right) \nonumber\\
 &+ \sum_{i=1}^{3} \frac{-2+\eta_{\phi^{i}}}{2}\left(\tilde{P}_{\phi, 1}
 \left( \frac{ \partial V}{\partial \phi^i} \frac{\partial V}{\partial \phi^i}\right)_{q, (r)}^{(x), (A)} - \tilde{P}_{\phi, 0}\left( \frac{ \partial^2 V}{\partial \phi^i\phi^i}\right)_{q, (r)}^{(x), (A)}
 \right)\nonumber\\
 & + \Biggl\lbrace\left(\frac{1}{3}-\frac{\eta_{v}}{2}\right)(2A_1 + A_4 + A_5 + r_1 +
 r_2)\nonumber\\&
 +\left(-D+\frac{1}{3}-\frac{\eta_{u}}{2}\right)(2A_2 + A_4 + A_6 + r_3 + r_4)\nonumber\\
 & + \left(-\frac{1}{3}-\frac{\eta_{f}}{2}\right)(2A_3 + A_5 + A_6 + r_5 + r_6) \nonumber\\ & +\left(-\frac{D}{2}+\frac{1}{3}-
\frac{\eta_{\Psi}}{2}\right)(r_1+r_2 +r_3+
 r_4 +r_5+ r_6 + 2q) \nonumber\\&+ 
 \sum_{i=1}^{3}\left(-\frac{D}{2}+1-\frac{\eta_{\phi^{i}}}{2}\right)x_i \left. -\frac{2}{3}\mathit{Der} +
 D \right\rbrace \nonumber\\
 &\times _{(x)}^{
 \;\;q}V_{(r)}^{(A)}[\mathit{Der}].
\end{align}
Here we defined
\begin{equation}
\left( \frac{ \partial V}{\partial v_i} \frac{\partial V}{\partial v_i}\right)_{q, (r)}^{(x), (A)}
\end{equation}
as the contribution of $\frac{ \partial V}{\partial v_i} \frac{\partial V}{\partial v_i}$ with the
indicated field expansion, and similarly for the other terms.

If the determinant (\ref{determinant}) is taken into account in a way that implies
additional fields, these have to be
included into the RG-equation in the same way.

In the theory described by the effective actions 
(\ref{wirkgeister}) or (\ref{locaction}) the 1-particle irreducible
Green functions of the velocity field $v$ vanish as a consequence of the fact
that in the effective action there is no $v$-propagator and there
are no vertices with the field $v$ only. On the other hand, integrating the
auxiliary fields out would produce an action containing a $v$-propagator and $v$-vertices,
leading to Green functions that are 1-particle irreducible within this theory.
Even though these properties have important consequences for
studies of the perturbation expansions of these actions, they don't
influence our numeric approach based on (\ref{rateeq}), as will become clear in a
subsequent paragraph.

Of course, it is a drawback to expand the action in powers of fields \emph{and}
momenta. For a numerical implementation of the RG flow, however, some
approximation scheme has to be chosen. This one enables us to work with a very
simple and fast numerical algorithm, which is described in the next section.

At this point, it is possible to examine the scaling of the two- and four-point
functions $\langle vv \rangle$
and $\langle vvvv \rangle$ near the free fixed point. The flow equations are
\begin{equation}
- \Lambda\frac{d}{d\Lambda}
\left\langle vv \right\rangle = -\frac{2}{3} \left\langle vv \right\rangle + 44 \tilde{P}_{v, 0}  \frac{\left\langle vvvv \right\rangle}{\left\langle vv \right\rangle}
\end{equation}
and
\begin{equation}
- \Lambda\frac{d}{d\Lambda}
\left\langle vvvv \right\rangle = -\frac{4}{3} \left\langle vvvv \right\rangle + 55 \tilde{P}_{v, 0} \frac{\left\langle vvvvvv \right\rangle}{\left\langle vv \right\rangle}.
\end{equation}
In the limit of small couplings, this reduces to
\begin{equation}
- \Lambda\frac{d}{d\Lambda}
\left\langle vv \right\rangle = -\frac{2}{3} \left\langle vv \right\rangle
\end{equation}
\begin{equation}
- \Lambda\frac{d}{d\Lambda}
\left\langle vvvv \right\rangle = -\frac{4}{3} \left\langle vvvv \right\rangle
\end{equation}
From this we obtain the scaling of the two- and four-point-function as
\begin{eqnarray}
\left\langle vv \right\rangle &\sim& (x)^\frac{2}{3},\\
\left\langle vvvv \right\rangle &\sim& (x)^\frac{4}{3}.
\end{eqnarray}
This is precisely the K41 scaling predicted by Kolmogorov.
$\tilde{P}_{v, 0}$ can be interpreted as a measure for the coupling, defining the
meaning of \emph{being near the free fixed point}.

\section{Numerical Analysis}

\subsection{Choice of Renormalisation Group Equation}

In developing the numerical algorithm, we tried different choices for the
cutoff, including the sharp cutoff of the Wegner-Houghton equation,
Eq.(~\ref{WH}). We found that this choice is particularly suitable, as it
allows us to compute the contribution for an infinitesimal integration step
independently of the coup\-lings involved.

The algorithm calculates the RG flow in terms of the coefficients
(\ref{coefficients}) of the derivative expansion. The flow equations are a
set of coupled ordinary differential equations for these coefficients, which
are solved numerically with given initial conditions. For the calculation of
the RG flow we worked with a predictor-corrector-, as well as a
Runge-Kutta-integration algorithm, both with self-adjusting step width. We
used two sets of algorithms - one of them involves explicitly programmed
versions of the rate equations, while the others worked out the loop- and
link-graphs automatically, only needing the parameters of the physical
system.

Apart from the algorithm for the calculation of the flow, we developed a
number of tools for the analysis of the resulting data. As the coupling
space, in which we are working, is very abstract and high-dimensional, it is
helpful to start with explorative studies of unphysical toy systems, i.e.\
simple and solvable physical systems, and of reduced turbulent systems
(Burgulence), to gain confidence in the correct working of the the
algorithm, and to develop some intuition for the work with renormalised
couplings.

\subsection{Non-Turbulent Systems}

We started our investigations by analysing unphysical (toy-)systems with
arbitrary constants and dimensionality of space, to learn more about the
detection and features of different sorts of fixed points. A main question
was how structures in coupling space can be recognised, if the
dimensionality of the coupling space is high, and whether terms of higher
order in the field expansion contribute as corrections.

In a second step, we applied the algorithm to physical systems with known
properties, such as the scalar and the O(3)-symmetric field theory, in order
to check that the algorithm works correctly and to see how closely we can
reproduce analytic values for fixed point scalings; and on the other hand to
approach turbulent hydrodynamics in a stepwise manner, interpreting it as a
special case of the general $3$-vector-model.

\subsubsection{Toy Systems} 

We worked with a number of unphysical systems for testing the algorithm and
analysis tools, thus merely looking for nontrivial structures. These systems
were defined by an action consisting of a propagator, a two-field- and a
four-field-interaction, where the field was a $3$-vector-field. Parameters
were deliberately adjusted to allow the presence of different fixed points.

\begin{figure}[H]
\begin{center}
\includegraphics[width=4in]{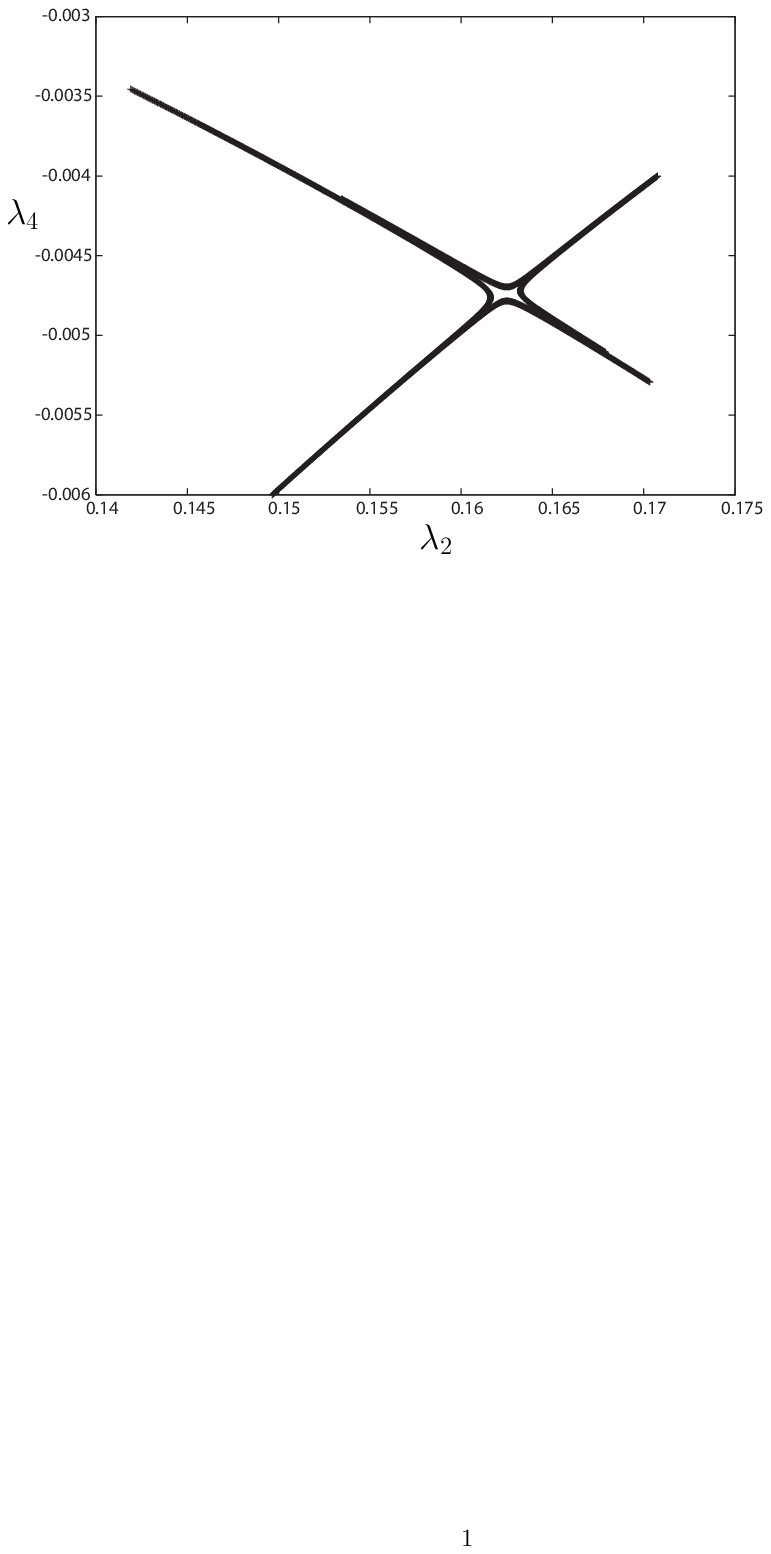}
\caption{\label{zoom}Fixed point of an unphysical model system, as found by
use of the shooting method. Shown is the flow in the two-field-interaction
$\lambda_2$ and the four-field-interaction $\lambda_4$.
The attractive direction goes to the upper right
and lower left corners of the diagram, the other two directions are
repulsive.}
\end{center}
\end{figure}

Investigations of the coupling space were mainly done using the shooting
method, which is especially useful for finding fixed points. In practice,
one initiates a number of RG-flows, starting from initial conditions
sufficiently close to each other, and searches the topology of the flow for
interesting structures. To identify the location of the fixed point more
precisely, one repeats the method with initial conditions closer to the
estimated fixed point couplings, leading to a picture like Fig.~\ref{zoom}.
In this way, one approaches the fixed point iteratively. Following this
procedure, the simulated trajectories approach the ideal trajectories, i.e.\
the flows directly running into our out of the fixed point.

The shooting method is limited by the numerical accuracy of the computer
program, and the stepsize adjustment of the flow integration, as the
algorithm slows down drastically when a fixed point is approached.

In a simulation involving more than two couplings, as is usually the case,
the projection of the flow onto a two-dimensional subspace will in general
not look so evident, but quite similar if the fixed point is approached
closely enough.

\subsubsection{Simple Physical Systems}

Using our algorithm, the renormalisation group flows of the scalar field
theory and the O(3)-symmetric theory in $D$ dimensions have been analysed by
P.~D\"uben \cite{dueben}. By reproducing known values of these theories like
fixed point locations and scaling (also in the $\epsilon$-expansion), we
went a step further towards the much more divert general
three-vector-theory, and again checked the correctness of our algorithms. We
found that we are able to accurately reproduce the values known from
literature, to a given order of the $\epsilon$-expansion. These results will
be published in a forthcoming article.

\subsection{Hydrodynamics near the Local Potential Approximation}

Now we return to the analysis of the action for hydrodynamics derived above
in the LPA. The system is specified by the dimensionality of space and
symmetries of the fields; the action in the LPA (\ref{locaction}) serves as
the initial condition of the flow.

In calculations of the RG flow it is generally preferable to calculate
$\eta$, rather than to search for it by means of the shooting method. In the
strict version of the LPA, on the other hand, one has $\eta = 0$ as no field
renormalisation is performed. We can extend the LPA by rescaling the field
such that the corresponding anomalous dimension $\eta$ equals some prescribed
value.

The calculations of the RG flow were performed using two distinct
algorithms: The first one iterating the rate equations derived in the
previous sections and doing the book-keeping of the terms involved
explicitly, the second one finding the graphs to be computed automatically.
The second formulation turned out to be not only more elegant, but a great
deal faster than the cumbersome implementation of the book-keeping.

The advantage of this approach is the fast integration of a large number of
couplings, and in that way evading the drawbacks of the expansions.
Calculations were done with up to 100 couplings, though it has to be said
that the identification of fixed points becomes nearly impossible in these
high-dimensional spaces. Working with such a number of terms can only be
done iteratively, meaning that one starts with a low number of couplings,
identifies the fixed point and than changes to more and more terms, hoping
that these act as corrections to the overall behaviour.

\begin{figure}[H]
\begin{center}
\includegraphics[width=4in]{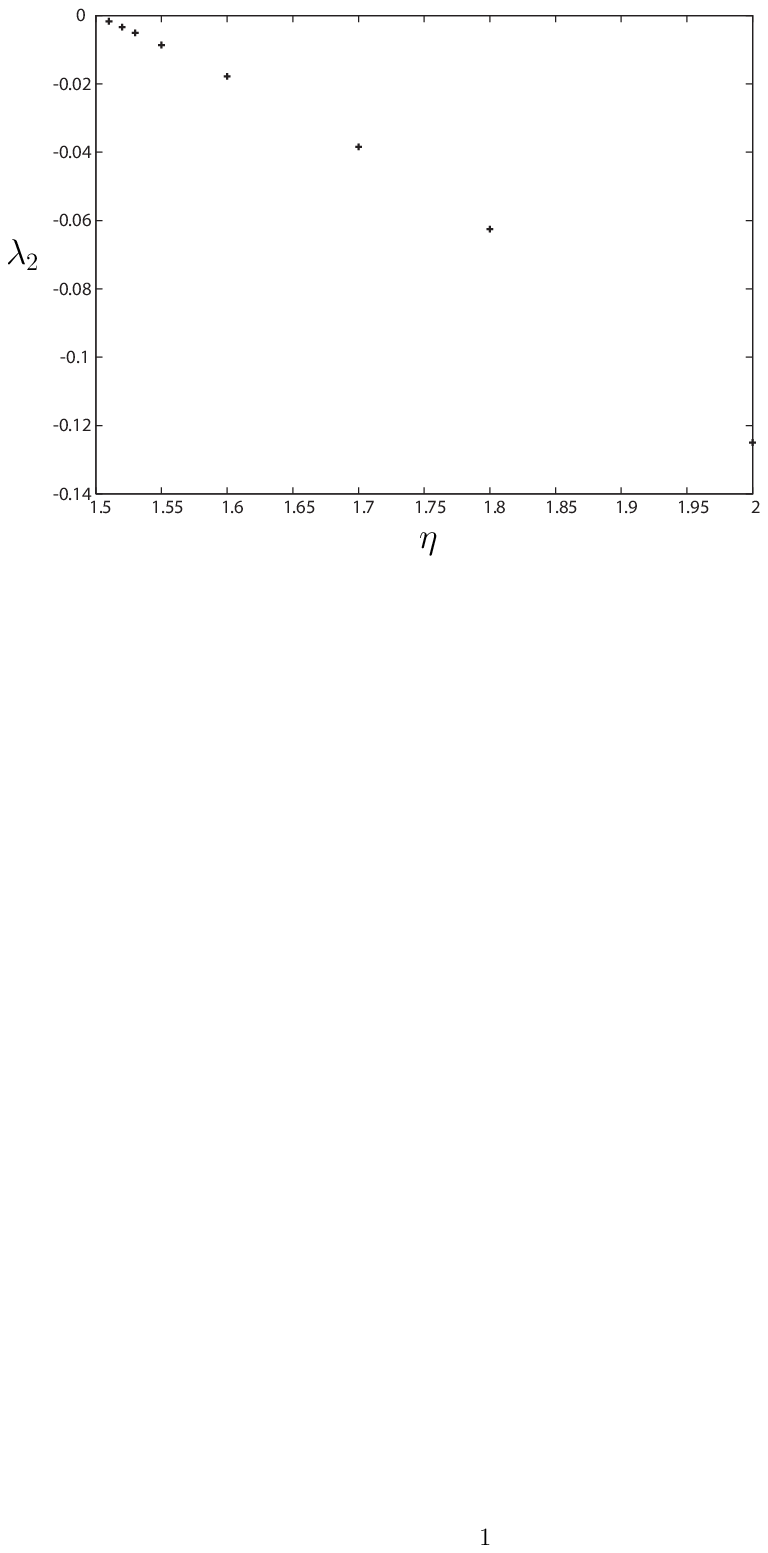}
\caption{\label{v2}Fixed point of hydrodynamics in the Local Potential
Approximation, as found by use of the shooting method. The plot shows the
fixed point value of the $\langle vv \rangle$-coupling $\lambda_2$, 
depending on the anomalous exponent $\eta$.}
\end{center}
\end{figure}

\begin{figure}[H]
\begin{center}
\includegraphics[width=4in]{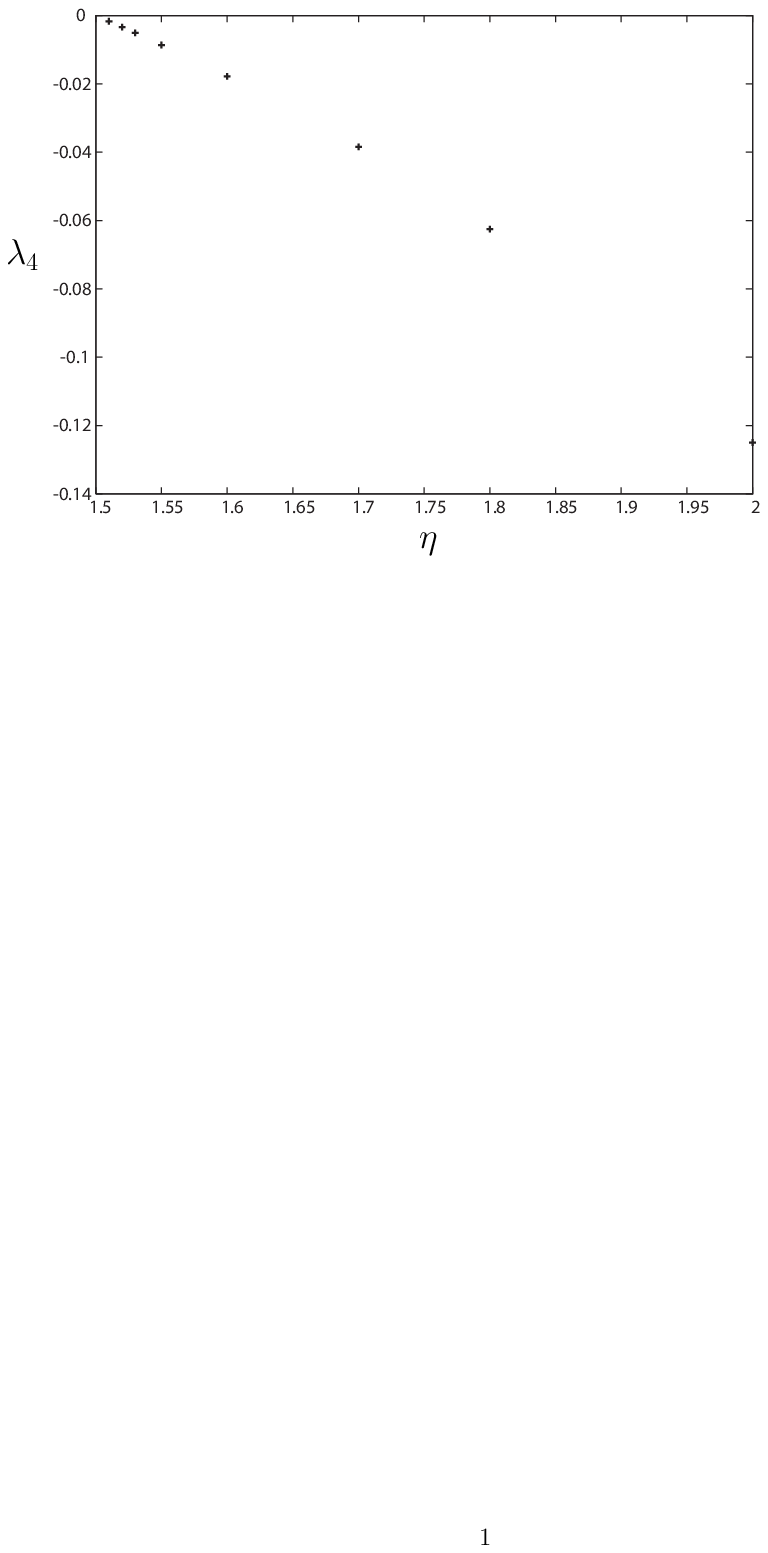}
\caption{\label{v4}Fixed point of hydrodynamics in the Local Potential
Approximation, as found by use of the shooting method. The plot shows the
fixed point value of the $\langle vvvv \rangle$-coupling $\lambda_4$, 
depending on the anomalous exponent $\eta$.}
\end{center}
\end{figure}

It is not difficult to show that for values $\eta > 1.5$ of the anomalous
exponent, a non-trivial fixed point exists in the vicinity of the trivial
one. We used the shooting method to determine the position of this
non-trivial fixed point, depending on the anomalous exponent, as can be seen
in Fig.~\ref{v2} and Fig.~\ref{v4}. The distance to the origin of coupling
space can be seen to grow linearly with $\eta$; we can, however, not relate
this fixed point to any physical property. For $\eta < 1.5$ this fixed point
does not exist.


\subsection{Scaling of the Trivial Fixed Point}

It is straightforward to analyse the scaling of the trivial fixed point. The
correlation functions of even orders are directly computed by the RG-flow;
after Fourier-transformation to physical space we can read off the scaling,
and find:
\begin{table}[H]
\centering
\begin{tabular}[h]{|c|c|}
\hline  Order of the Correlation Function & Scaling Exponent \\ \hline 
\hline 
 2 & $0,666 \pm 0,017$ \\
 \hline
4 & $ 1,338 \pm 0,035$ \\
\hline 6 & $ 1,999 \pm 0,052$ \\
\hline 
\end{tabular}
\caption{Scaling exponents at the trivial fixed point}
\end{table}
The correlation functions of odd orders are not explicit terms of the action
and so have to be measured indirectly. The correlation function of order $n$
can, for example, be derived from the term $\langle uv^n \rangle$, if the
scaling of the field $u$ is known. We chose to measure the scaling of $u$
from the two-point-function $\langle uu \rangle$, and subtract it from
$\langle uv^n \rangle$. In this way the following exponents can be measured:
\begin{table}[H]
\centering
\begin{tabular}[h]{|c|c|}
\hline  Order of the Correlation Function & Scaling Exponent \\ \hline 
\hline 
 1 & $0,3334 \pm 0,0018$ \\
 \hline
3 & $ 1,0004 \pm 0,0012$ \\
\hline 
5 & $ 1,6681 \pm 0,0012$ \\
\hline
7 & $ 2,3348 \pm 0,0012$ \\
\hline 
\end{tabular}
\caption{Scaling exponents at the trivial fixed point}
\end{table}
\noindent
These numbers demonstrate that the trivial fixed point represents the
scaling of Kolmogorov's K41-theory.

\section{Conclusions and Outlook}

We have shown how to define a generating functional for hydrodynamic
turbulence, including a strict treatment of the incompressibility condition.
The non-local interactions have been transformed into local ones by means of
auxiliary fields. In addition, we have applied a derivative expansion to
approximate the resulting action.

Concerning the renormalisation group, we discussed the procedure of
renormalisation and rescaling in some detail. We obtained a RG-equation for
a general multi-component action, including the turbulent action, and a set
of rate equations after application of the derivative expansion.

Our numerical algorithm allows to compute the RG flow in this setting,
including products of Grassmannian variables. We tested the numerical
algorithm by reproducing known values for non-trivial scalings of the scalar
theory in $4-\epsilon$ dimensions, and the O(3)-symmetric field theory. The
results are in agreement with values found in literature, giving us
confidence in the reliability of the numerical algorithm.

In the context of turbulence we were able to identify the trivial
fixed-point with the scaling exponents predicted by the K41-theory.

So far we have not been able to reproduce the intermittent exponents for the
structure functions of fully developed turbulence that would agree with the
experimental values. The reason for this deficit lies in the complexity of
the general 3-vector-model, including all theories that are based on
hydrodynamics. Although the basic foundations of these theories are well
understood, all of them (including Navier-Stokes and Burgers turbulence)
involve the same dimensionality of space and symmetry of the fields, while
leading to different predictions for the intermittent exponents.

Finally, it should be noted that it is not clear whether the analysis of a
fixed point will eventually lead to an understanding of intermittency.
Available data on turbulence show that the probability distribution of the
velocity increment looks, for small distances, like a L\'{e}vy-distribution;
on large scales like normally distributed \cite{friedrich}. This could be an
indication for a crossover between two fixed points. It would be interesting
to test this conjecture by future flow calculations with our algorithm.

\bibliographystyle{plain}

\begin{thebibliography}{10}

\bibitem{aav}
L.T.~Adzhemyan, N.V.~Antonov and A.N.~Vasiliev.
\newblock {\em The Field Theoretic Renormalization Group in Fully Developed
Turbulence}.
\newblock London, 1999.

\bibitem{ballthorne}
R.D.~Ball and R.S.~Thorne.
\newblock {\em Annals Phys.}, 236:117, 1994.

\bibitem{bec}
J.~Bec and K.~Khanin.
\newblock arXiv:0704.1611, 2007.

\bibitem{bg}
G.~Benfatto and G.~Gallavotti.
\newblock {\em Renormalization Group}.
\newblock Princeton, 1995.

\bibitem{bh2007}
A.~Berera and D.~Hochberg.
\newblock {\em Phys.~Rev.~Lett.}, 99:254501, 2007.

\bibitem{bervillier}
C.~Bervillier.
\newblock {\em Phys.~Lett.~A}, 332:93, 2004.

\bibitem{callan}
C.G.~Callan.
\newblock {\em Phys.~Rev.~D}, 2:1541, 1970.

\bibitem{coltom}
R.~Collina and P.~Tomassini.
\newblock {\em Phys.~Lett.~B}, 411:117, 1997.

\bibitem{collins}
J.~Collins.
\newblock {\em Renormalization}.
\newblock Cambridge, 1984.

\bibitem{dedommartin}
C.~De Dominicis and P.C.~Martin.
\newblock {\em Phys.~Rev.~A}, 19:419, 1979.

\bibitem{dueben}
P.~D\"uben.
\newblock {\em Numerische Anwendungen des Pfadintegralformalismus in
hydrodynamischer Turbulenz}.
\newblock Diploma thesis, Univ.~of M\"unster, 2009.

\bibitem{epl}
P.~D\"uben et~al.
\newblock {\em Europhys.~Lett.}, 84:40002, 2008.

\bibitem{eg}
A.~Esser and S.~Gro{\ss}mann.
\newblock{\em Phys.~J.~B} 7:467, 1999.

\bibitem{fns}
D.~Forster, D.R.~Nelson and M.J.~Stephen.
\newblock {\em Phys.~Rev.~A}, 16:732, 1977.

\bibitem{friedrich}
R.~Friedrich.
\newblock {\em Phys.~Rev.~Lett.}, 90:084501, 2003.

\bibitem{frisch}
U.~Frisch.
\newblock {\em Turbulence}.
\newblock Cambridge, 1995.

\bibitem{frischbec}
U.~Frisch and J.~Bec.
\newblock Burgulence.
\newblock In A.Yaglom M.~Lesieur and F.~David, editors, {\em Les Houches 2000:
New Trends in Turbulence}, pages 341--383. Springer EDP-Sciences, 2001.

\bibitem{golner}
G.R.~Golner.
\newblock hep-th/9801124v3, 2000.

\bibitem{hasenfratz}
A.~Hasenfratz and P.~Hasenfratz.
\newblock {\em Nucl.~Phys.~B}, 270:687, 1986.

\bibitem{mydiss}
D.~Homeier.
\newblock {\em Renormierungsgruppenflussgleichungen und hydrodynamische
Turbulenz}.
\newblock Ph.D.~thesis, Univ.~of M\"unster, 2006.

\bibitem{kadanoff}
L.P.~Kadanoff.
\newblock {\em Physics}, 2:263, 1966.

\bibitem{kolmogorov}
A.N.~Kolmogorov.
\newblock {\em Dokl.~Akad.~Nauk SSSR}, 32:16, 1941.

\bibitem{ll}
L.D.~Landau and E.M.~Lifshitz.
\newblock {\em Course of Theoretical Physics, Vol.~6, Fluid Mechanics}.
\newblock Oxford, 1987.

\bibitem{laxlev}
P.D.~Lax and C.D.~Levermore.
\newblock {\em Commun.~Pure Appl.~Math.}, 36:253,571,809, 1983.

\bibitem{lvovproc}
V.~L'vov and I.~Procaccia.
\newblock {\em Exact Resummations in the Theory of Hydrodynamic Turbulence}.
\newblock Les Houches, 1994.

\bibitem{msr}
P.C.~Martin, E.D.~Siggia and H.A.~Rose.
\newblock {\em Phys.~Rev.~A}, 8:423, 1973.

\bibitem{mccomb}
W.~McComb.
\newblock {\em The Physics of Fluid Turbulence}.
\newblock Oxford, 1990.

\bibitem{my}
S.~Monin and A.M.~Yaglom.
\newblock {\em Statistical Fluid Mechanics}.
\newblock Moskau, 1965.

\bibitem{morris}
T.R.~Morris.
\newblock {\em Prog.~Theor.~Phys.~Suppl.}, 131:395, 1998.

\bibitem{munoz}
G.~Mu\~{n}oz and W.S.~Burgett.
\newblock {\em Journal of Statistical Physics}, 56:59, 1989.

\bibitem{pesschroe}
M.~Peskin and D.~Schroeder.
\newblock {\em An Introduction to Quantum Field Theory}.
\newblock Cambridge, 1995.

\bibitem{pope}
S.B.~Pope.
\newblock {\em Turbulent Flows}.
\newblock Cambridge, 2000.

\bibitem{sornette}
D.~Sornette.
\newblock {\em Phys.~Rep.}, 297:239, 1998.

\bibitem{symanzik}
K.~Symanzik.
\newblock {\em Commun.~Math.~Phys.}, 18:227, 1970.

\bibitem{wegnerhoughton}
F.~Wegner and A.~Houghton.
\newblock {\em Phys.~Rev.~A}, 8:401, 1973.

\bibitem{wilson}
K.G.~Wilson.
\newblock {\em Phys.~Rev.~B}, 4:3174, 1971.

\bibitem{wilsona}
K.G.~Wilson.
\newblock {\em Phys.~Rev.~B}, 4:3184, 1971.

\bibitem{wk}
K.G.~Wilson and J.~Kogut.
\newblock {\em Phys.~Rep.~C}, 12:75, 1974.

\bibitem{zj}
J.~Zinn-Justin.
\newblock {\em Quantum Field Theory and Critical Phenomena (4th ed.)}.
\newblock Oxford, 2002.

\end{thebibliography}

\end{document}